%% file: sb-21-submit.tex
\documentclass[12pt]{article}
\usepackage{color,amsmath,amssymb,graphicx}
\definecolor{refkey}{rgb}{1,0.5,0.5}
\definecolor{labelkey}{rgb}{0.5,1,0.5}
\include{abbrevs}
 \numberwithin{equation}{section}

\begin{document} 
\title{Symmetry breaking bifurcation in Nonlinear Schr\"odinger  /Gross-Pitaevskii Equations}
\author{E.W. Kirr%
\thanks{Department of Mathematics, University of Ilinois, Urbana-Champaign}\ , \
P.G. Kevrekidis
\thanks{Department of  Mathematics and Statistics,
University of Massachusetts, Amherst, MA 01003}\ ,\
E. Shlizerman
\thanks{Department of Computer Science and Applied Mathematics,
Weizmann Institute of Science, Rehovot, Israel}\ ,\  and
M.I. Weinstein
\thanks{Department of Applied Physics and Applied Mathematics,
Columbia University, New York, NY 10027}
}
\maketitle
\begin{abstract}
We consider a class of nonlinear Schr\"odinger / Gross-Pitaveskii  (NLS-GP) equations, {\it i.e.}  NLS with a linear potential. We obtain conditions for a symmetry breaking bifurcation in a symmetric family of states as $\cN$, the squared $L^2$ norm 
 (particle number, optical power), is increased.  In the special case where the  linear potential is a double-well with well separation $L$,  we estimate $\cN_{cr}(L)$, the symmetry breaking threshold.
 Along the ``lowest energy'' symmetric branch, there is an exchange of stability from the symmetric to asymmetric branch as $\cN$ is increased beyond $\cN_{cr}$.
  \end{abstract}
  \tableofcontents
\section{Introduction}\label{sec:introduction}
Symmetry breaking is a ubiquitous and important phenomenon which arises in a wide range of physical systems. In this paper, we consider a class PDEs, which are invariant under a symmetry group.  For sufficiently small values of a parameter, $\cN$, the preferred (dynamically stable) stationary (bound) state of the system is invariant under this symmetry group. However, above a critical parameter, $\cN_{cr}$, although the group-invariant state persists, the preferred state of the system is a state which (i) exists only for $\cN>\cN_{cr}$ and (ii) is no longer invariant. That is,  symmetry is broken and there is an exchange of stability.

Physical examples of symmetry breaking include
liquid crystals \cite{vps:98}, quantum
dots \cite{YL:99}, semiconductor lasers \cite{HFE:01} and pattern
dynamics \cite{SMS:00}.
 This article focuses on  spontaneous symmetry breaking, as a phenomenon in nonlinear optics
\cite{CSMVKEH:02,KCMFW:05,KapitulaKevrekidesChen},
as well as in  the
macroscopic quantum setting of Bose-Einstein condensation (BEC)
  \cite{AGFHCO:05}.  Here, the governing equations are
  partial differential equations (PDEs) of nonlinear Schr\"odinger / Gross-Pitaevskii type (NLS-GP).
 Symmetry breaking has been observed experimentally in optics
for two-component
spatial optical vector solitons (i.e., for self-guided laser beams in
Kerr media and focusing cubic nonlinearities) in \cite{CSMVKEH:02},
as well as
for the electric field distribution between two-wells of a
photorefractive crystal in \cite{KCMFW:05} (and between three
such wells in \cite{KapitulaKevrekidesChen}). In BECs,
an effective double well formed by a combined (parabolic) magnetic
trapping and a (periodic) optical trapping of the atoms may have
similar effects \cite{AGFHCO:05}, and lead to ``macroscopic quantum self-trapping''.

Symmetry breaking in ground states of the three-dimensional NLS-GP equation, with an attractive nonlinearity of Hartree-type and a symmetric double well linear potential, was considered in Aschbacher {\it et. al.}
 \cite{AFGST:02}; see also Remark \ref{rmk:variational-groundstate}.   {\it Ground states} are positive and symmetric nonlinear bound states, arising as  {\it minimizers} of, $\cH$,  the NLS-GP Hamiltonian energy subject to fixed, $\cN$, the squared $L^2$ norm. For the class of equations considered in \cite{AFGST:02}, ground states exist for any $\cN>0$. It is  proved that for sufficiently large $\cN$, any ground state is concentrated in only one of the wells, {\it i.e.} symmetry is broken. The
  analysis in \cite{AFGST:02}  is an asymptotic study for large $\cN$,
showing that if $\cN$ is sufficiently large, then it is energetically
preferable for the ground state to localize in a single well.
 In contrast,  at small
$L^2$ norm the ground state is bi-modal, having the symmetries of the   linear Schr\"odinger operator with symmetric double-well potential.
 For macroscopic quantum systems, the squared $L^2$ norm, denoted
by $\cN$, is the particle number, while in optics it is the optical power.
 An attractive nonlinearity corresponds to the case of negative scattering length in BEC and positive attractive Kerr nonlinearity in optics.

An alternative approach to symmetry breaking in NLS-GP is via bifurcation theory. It follows from \cite{RW:88,pw:97} that a family of ``nonlinear ground states'' bifurcates from the zero solution ($\cN=0$) at the ground state energy of the Schr\"odinger operator with a  linear double well potential. This nonlinear ground state branch consists of states having the same bi-modal symmetry of the linear ground state. In this article we prove, under suitable conditions, that  there is a secondary bifurcation to an asymmetric state at critical $\cN=\cN_{cr}>0$. Moreover, we show that there is a transfer or exchange of stability which takes place at $\cN_{cr}$; for $\cN<\cN_{cr}$ the symmetric state is stable, while for $\cN>\cN_{cr}$ the asymmetric state is stable.
 Since our method is based on local bifurcation analysis we do not require that the states we consider satisfy a minimization principle, as in \cite{AFGST:02}. Thus, quite generally, symmetry-breaking occurs  as a consequence of the (finite dimensional) {\it normal form}, arising in systems with certain symmetry properties.  Although we can treat a large class of problems for which there is no minimization principle, our analysis, at present, is restricted to small norm. As we shall see, this can be ensured, for example,  by taking the distance between wells in the double-well, to be sufficiently large.

  In \cite{JW:04} the precise transition point to symmetry breaking,
$\cN_{cr}$, of the ground state and the transfer of its stability to an
asymmetric ground state was considered (by geometric dynamical systems methods) in the  exactly solvable NLS-GP, with a double well potential consisting of two Dirac delta functions, separated by a
distance $L$.
  Additionally, the behavior of the function $\cN_{cr}(L)$, was considered.
Another solvable model was examined by numerical means
in \cite{MKR:02}. A study of dynamics for nonlinear double wells
appeared in \cite{S:03}.

  We study $\cN_{cr}(L)$, in general.
$\cN_{cr}(L)$, the value  at which symmetry breaking occurs, is closely related to the
spectral properties of the linearization
   of NLS-GP about the symmetric branch. Indeed, so long as the linearization
of NLS-GP at the symmetric state has no non-symmetric null space,
the symmetric state is locally unique, by the implicit function theorem \cite{Nirenberg}.
The mechanism for symmetry breaking is the first appearance of
an anti-symmetric element in the null space of the linearization
for some $\cN=\cN_{cr}$. This is demonstrated for a finite dimensional
Galerkin approximation of NLS-GP in \cite{KCMFW:05,KK:05}.
The present work extends and generalizes this analysis to the full
infinite dimensional problem using the Lyapunov-Schmidt method \cite{Nirenberg}. Control of the corrections to the finite-dimensional approximation requires small norm of the states considered. Since, as anticipated by the Galerkin approximation, $\cN_{cr}$ is proportional to the distance between the lowest eigenvalues of the double well,
 which is exponentially small in $L$, our results apply to double wells with separation $L$,  hold for $L$ sufficiently large.

\begin{figure}[ht]
\begin{center}
\includegraphics[height=6cm]{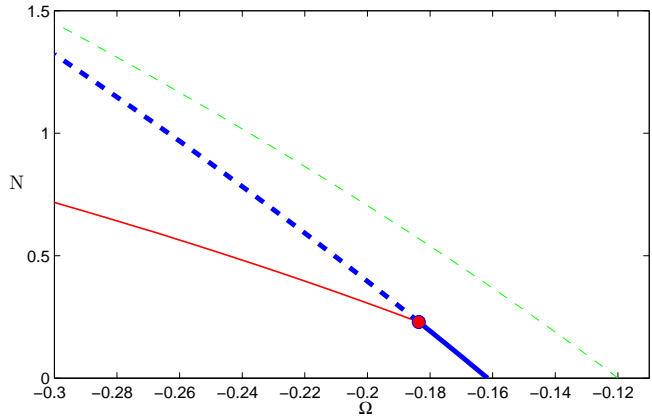}
\end{center}
\caption{(Color Online) Bifurcation diagram for NLS-GP with double well potential (\ref{poten}) with parameters $s=1 , L=6$ and cubic nonlinearity.
The first bifurcation is from the the zero state at the {\it ground state energy} of the double well.
 Secondary bifurcation to an asymmetric state at $\cN=\cN_{cr}$ is marked by
a (red) circle. For $\cN < \cN_{cr}$ the symmetric state
(thick (blue) solid line) is nonlinearly dynamically stable. For $\cN > \cN_{cr}$
the symmetric state is unstable  (thick (blue) dashed line). The
stable asymmetric state, appearing for $\cN
> \cN_{cr}$, is marked by a thin (red) solid line. The (unstable)
antisymmetric state is marked by a thin (green) dashed
line.} \label{fig:gen_bifdiag}
\end{figure}

The article is organized as follows. In section \ref{sec:tformulation}
 we introduce the NLS-GP model and give a technical formulation of the
bifurcation problem. In section \ref{sec:finite-dim-approx} we study a
finite dimensional truncation of the bifurcation problem, identifying
a relevant bifurcation
point. In section \ref{sec:bifsymbrk}, we prove the persistence of  this
symmetry breaking bifurcation in the full NLS-GP problem,
for $\cN\ge\cN_{cr}$. Moreover, we show that the lowest energy symmetric state becomes
dynamically unstable at $\cN_{cr}$ and the bifurcating asymmetric state is the
dynamically stable ground state for $\cN>\cN_{cr}$.
Figure \ref{fig:gen_bifdiag} shows a typical bifurcation diagram demonstrating  symmetry breaking for
the NLS-GP system with a double well potential. At the
bifurcation point $\cN_{cr}$ (marked by a circle in the figure), the symmetric  ground state becomes unstable and a stable asymmetric state emanates from it.

 The main results are stated in Theorem \ref{theo:symbrkbif}, Corollary \ref{cor:dwell-sb} and Theorem \ref{theo:exchange-stab}. In particular, we obtain an asymptotic formula for the critical particle number (optical power) for symmetry breaking in NLS-GP,
  \begin{equation}
\cN_{cr}\ =\ \frac{\Omega_1-\Omega_0}{\Xi[\psi_0,\psi_1]}
\ +\
  \cO\left(\ \frac{ (\Omega_1-\Omega_0)^{2} }{ \Xi[\psi_0,\psi_1]^{3} }\right).
  \label{Ncr-approx}
  \end{equation}
   Here, $(\Omega_0,\psi_0)$ and $(\Omega_1,\psi_1)$ are eigenvalue - eigenfunction pairs of the {\it linear} Schr\"odinger operator $H=-\Delta+V$, where $\Omega_0$ and $\Omega_1$ are separated from other spectrum, and $\Xi$ is a positive constant, given by (\ref{Xi-def}), depending on $\psi_0$ and $\psi_1$. {\it  The most important case is where $\Omega_0 < \Omega_1$ are the \underline{lowest} two energies (linear ground and first excited states).} For double wells with separation $L$, we have $\cN_{cr}=\cN_{cr}(L)$, depending on the eigenvalue spacing $\Omega_0(L)-\Omega_1(L)$, which  is exponentially small if $L$ is large and $\Xi$ is of order one. Thus,  for large $L$,  the bifurcation occurs at small $L^2$ norm.
This is the weakly nonlinear regime in which the corrections to the finite
dimensional model can be controlled perturbatively. A local bifurcation diagram of this type will occur for {\it any} simple even-odd symmetric pair of simple eigenvalues of $H$ in the weakly nonlinear regime, so long as the eigen-frequencies are separated from the rest of the spectrum of $H$; see Proposition \ref{prop:eta} and the {\bf Gap Condition} (\ref{gapcond}). Therefore, a similar phenomenon occurs for  higher order, nearly degenerate pairs of eigen-states of the double wells, arising from isolated single wells with multiple eigenstates.
  Section \ref{sec:numerics} contains numerical results validating our
theoretical analysis.
\bigskip

\nit{\bf Acknowledgements:} The authors acknowledge the support of the US National Science Foundation, Division of Mathematical Sciences  (DMS).
 EK was partially supported by grants DMS-0405921 and DMS-060372. PGK was  supported, in part, by DMS-0204585,
NSF-CAREER and DMS-0505663, and acknowledges valuable discussions
with T. Kapitula and Z. Chen. MIW was supported, in part, by
DMS-0412305 and DMS-0530853. Part of this research was done while
Eli Shlizerman was a visiting graduate student in the Department of
Applied Physics and Applied Mathematics at Columbia University.
\section{Technical formulation}\label{sec:tformulation}

Consider the initial-value-problem for the time-dependent nonlinear Schr\"odinger / Gross-Pitaevskii equation (NLS-GP)
\begin{align}
i\D_t\psi\ &=\ H\psi\ +\ g(x) K[\ \psi\bar\psi\ ]\ \psi,\ \ \ \psi(x,0)\ {\rm given}
\label{nls-gp}\\
H\ &=\ -\Delta + V(x).\label{H-def}
\end{align}
We assume:
\begin{itemize}
\item[{\bf (H1)}]\  The initial value problem for NLS-GP is well-posed in the space $C^0([0,\infty);H^1(\R^n))$.
\end{itemize}

\nit {\bf (H2)}\ The  potential, $V(x)$ is assumed to be  real-valued , smooth and rapidly decaying as $|x|\to\infty$. The basic example of $V(x)$, we have in mind is a double-well potential, consisting of two identical potential wells, separated by a distance $L$.   Thus, we also assume symmetry with respect to the hyperplane, which without loss of generality can be taken to be $\{x_1=0\}$:
\begin{equation}
V(x_1,x_2,\dots,x_n)\ =\  V(-x_1,x_2,\dots,x_n).
\label{Vsym}
\end{equation}
 We assume the nonlinear term, $K[\psi\overline{\psi}]$,  to be attractive, cubic ( local or nonlocal), and symmetric in one variable. Specifically, we assume the following\\
\nit{\bf (H3)\ Hypotheses on the nonlinear term:}\
\begin{itemize}
\item[(a)]\ $g(x_1,x_2,\ldots ,x_n)=g(-x_1,x_2,\ldots ,x_n)$ (symmetry)
\item[(b)]\ $g(x) < 0$ (attractive\ /\ focusing)
\item[(c)]\ $K[h]=\int K(x-y)h(y)dy,\ \ \ K(x_1,x_2,\ldots ,x_n)=K(-x_1,x_2,\ldots ,x_n),\ \ \ K>0$.
\item[(d)]\
Consider the map $ N:H^2\times H^2\times H^2\mapsto L^2$ defined by
\begin{equation}
N(\phi_1,\phi_2,\phi_3)=gK[\phi_1\phi_2]\phi_3.
\label{Nphi123def}
\end{equation}
We also write $N(u)=N(u,u,u)$ and note that $\D_uN(u)=N(\cdot,u,u)+N(u,\cdot,u)+N(u,u,\cdot)$.
  We assume
  there exists a
 constant $k>0$ such that
 \be\label{N-est0} \|N(\phi_0,\phi_1,\phi_2)\|_{L^2}\ \le\
 k\ \|\phi_1\|_{H^2}\|\phi_2\|_{H^2}\|\phi_3\|_{H^2}.
 \ee
 \end{itemize}
Several illustrative and important examples are now given:

\nit{\bf Example 1:} Gross-Pitaevskii equation for BECs
with negative scattering\\
length $g(x)\equiv -1$, $K(x)=\delta(x)$\\
\nit{\bf Example 2:} Nonlinear Schr{\"o}dinger equation for
optical media with a nonlocal kernel $g(x) \equiv \pm 1$,
$K(x)=A \exp(-x^2/\sigma^2)$ \cite{krolikowski:04}
(see also \cite{deconinck:03} for similar considerations in BECs).\\
\nit{\bf Example 3:} Photorefractive nonlinearities
 The  approach  of the current paper can be adapted to the setting of photorefractive
crystals with saturable nonlinearities and appropriate optically
induced potentials \cite{moti05}. The
relevant symmetry breaking phenomenology is experimentally observable,
as shown in \cite{KCMFW:05}.
\bigskip

 \nit{\bf Nonlinear bound states:}
Nonlinear bound states are solutions of NLS-GP of the form
\begin{equation}
\psi(x,t)\ =\ e^{-i\Omega t} \Psi_\Omega(x), \label{boundstate}
\end{equation}
where $\Psi_\Omega\in H^1(\R^n)$ solves
\begin{equation}
H\ \Psi_\Omega\ +\ g(x)\ K[|\Psi_\Omega|^2]\ \Psi_\E\ -\ \E \Psi_\E\ =\ 0,\ \
 u\in H^1
\label{bs-eqn}
\end{equation}

If the potential $V(x)$ is such that the
operator $H=-\Delta+V(x)$ has a discrete eigenvalue, $E_* $, and correspsonding eigenstate $\psi_*$,  then for energies near $E$ near $E_*$ and one  expects small amplitude {\it nonlinear} bound states,  which are to leading order small multiples of $\psi_*$.  This is the standard setting of bifurcation from a simple eigenvalue \cite{Nirenberg}, which follows from the implicit function theorem.
\begin{theo}\label{theo:RW-theorem}\cite{pw:97, RW:88}
Let $(\Psi,E)=(\psi_*,E_*)$ be a  simple eigenpair,
of the eigenvalue problem $H\Psi=\E\Psi$, {\it i.e.} $dim\{\rho: (H-E_*)\rho=0\}=1$. Then,  there exists a unique smooth curve of nontrivial solutions $\alpha\mapsto (\ \Psi(\cdot;\alpha),\E(\alpha)\ )$, defined in a neighborhood of $\alpha=0$, such that
\begin{equation}
\Psi_\Omega\ =\ \alpha\left(\ \psi_0\ +\ \cO(|\alpha|^2)\ \right),\ \
 \Omega\ =\ \Omega_0\ +\ \cO(|\alpha|^2),\ \ \alpha\to0.
 \label{rw-states}
 \end{equation}
\end{theo}
\begin{rmk}\label{rmk:variational-groundstate}
For a large class of problems, a nonlinear ground state can be characterized variationally as a constrained minimum of the NLS / GP energy subject to fixed squared $L^2$ norm.
Define the NLS-GP Hamiltonian energy functional
\begin{equation}
\cH_{NLS-GP}[\Phi]\ \equiv\  \int |\nabla\Phi|^2\ +\ V|\Phi|^2\ dy\ +\ {1\over2}\ \int\ g(y){\cal K}[|\Phi|^2]\ dy\label{cH-NLSGP}
\end{equation}
and the particle number (optical power)
\begin{equation}
\cN[\Phi]\ =\ \int\ |\Phi|^2\ dy,
\label{Nconstraint}
\end{equation}
where
\begin{equation}
\cK[|\Phi|^2]\ =\ \int\ K(x-y)\ |\Phi(x)|^2\ |\Phi(y)|^2\ dy.
\label{K-def}\end{equation}
In particular, the following can be proved:
\begin{theo}\label{variational-gs}
Let $I_\lambda\ =\ \inf_{\cN[f]=\lambda}\ \cH[f]$. If
 $-\infty<I_\lambda<0$, then the minimum is attained at a positive solution of (\ref{bs-eqn}). Here, $\Omega=\Omega(\lambda)$ is a Lagrange multiplier for the constrained variational problem.
\end{theo}
   In \cite{AFGST:02} the nonlinear Hartree equation is studied;
   $\cK[h]=|y|^{-1}\star h$, $g\equiv -1$.  It is  proved that  if $V(x)$  is a double-well potential, then for $\lambda$ sufficiently large, the minimizer does not have the same symmetry as the linear ground state.
   By uniqueness, ensured by the implicit function theorem, for small $\cN$, the minimizer has the same symmetry as that as the linear ground state and has the expansion (\ref{rw-states}); see \cite{AFGST:02} and section \ref{sec:bifsymbrk}.
\end{rmk}

We make the following
\medskip

\nit{\bf Spectral assumptions on $H$}
\begin{itemize}
\item[{\bf (H4)}]\ $H$ has a pair of simple eigenvalues $\Omega_0$ and
$\Omega_1$.  $\psi_0$ and $\psi_1$, the corresponding (real-valued)
eigenfunctions are, respectively, even and odd in $x_1.$
\end{itemize}
\bigskip

\begin{example}\label{example:basic-example}\ {\bf The basic example:}  {\it Double well potentials}

\nit A class of examples of great interest is that of
double well potentials. The simplest example, in one space dimension,  is obtained as follows; see section \ref{sec:doublewells} for the multidimensional case.  Start with a single potential well (rapidly decaying as $|x|\to\infty$), $v_0(x)$, having exactly one eigenvalue, $\omega$,
 $H_0\psi_\omega=(-\Delta+v_0(x))\psi_\omega=\omega\psi_\omega$; see Figure (\ref{fig:sindoubwell}a). Center this
well at $x=-L$ and place an identical well, centered at $x=L$. Denote by $V_L(x)$ the resulting {\it double-well
potential} and $H_L$ denote the Schr\"odinger operator:
 \begin{equation}
 H_L=-\Delta+V_L(x)
 \label{HL-def}
 \end{equation}
\nit There exists  $L>L_0$, such that for $L>L_0$, $H_L$ has a pair of eigenvalues, $\Omega_0=\Omega_0(L)$ and
$\Omega_1=\Omega_1(L)$, $\Omega_0<\Omega_1$,  and corresponding eigenfunctions $\psi_0$ and $\psi_1$; see Figure
(\ref{fig:sindoubwell}b). $\psi_0$ is symmetric with respect to $x=0$ and
 $\psi_1$ is  antisymmetric with respect to $x=0$. Moreover, for $L$ sufficiently large, $|\Omega_0-\Omega_1| = \cO(e^{-\kappa L}),\ \kappa>0$; see \cite{Harrell}; see also section \ref{sec:doublewells}.

\begin{figure}[ht]
\begin{center}
\begin{tabular}{cc}
(a) & (b) \\
\includegraphics[height=8cm]{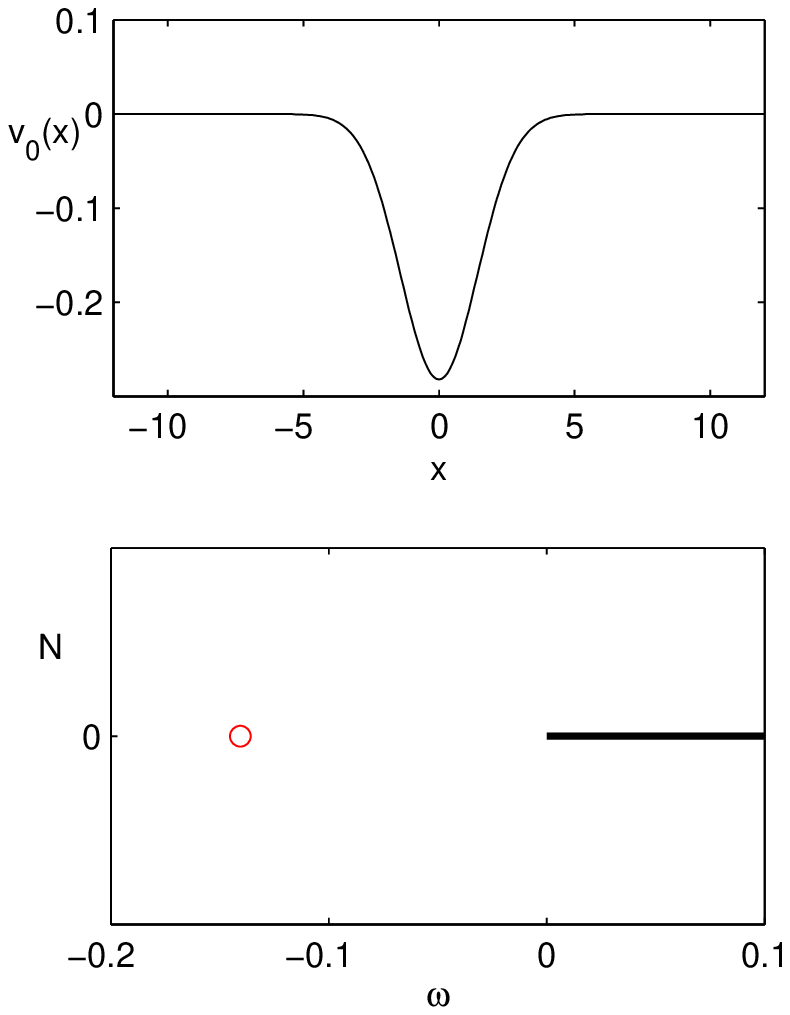} &
\includegraphics[height=8cm]{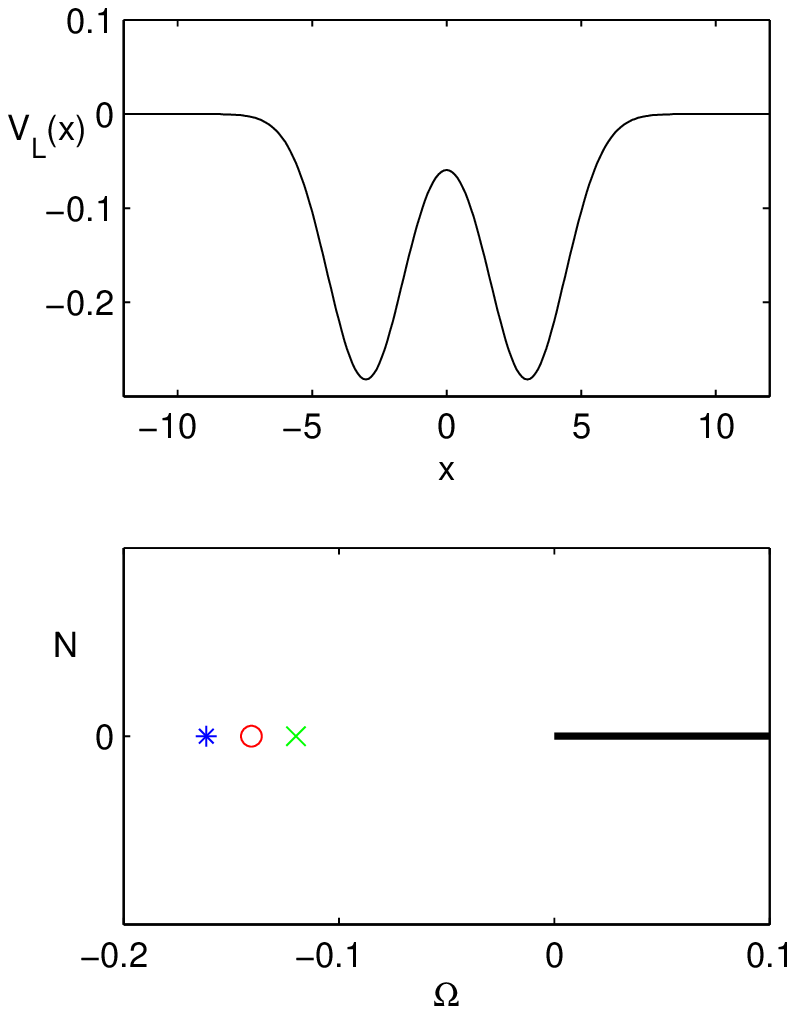} \\
\end{tabular}
\end{center}
\caption{This figure demonstrates a single
and a double well potential and the spectrum of
$H$ and $H_L$ respectively. Panel (a) shows a single well potential and under it the spectrum of $H$, with an eigenvalue
marked by a (red) mark `o' at $\omega$ and continuous spectrum marked by a (black) line for energies $\omega\ge0$. Panel
(b) shows the double well centered at $\pm L$ and the spectrum of $H_L$ underneath. The eigenvalues $\Omega_0$ and
$\Omega_1$ are each marked by a (blue) mark `*' and a (green) mark `x' respectively on either side of the location
$\omega$ - (red) mark `o'. The continuous spectrum is marked by a (black) line for energies $\Omega\ge0$. }
\label{fig:sindoubwell}
\end{figure}

The construction can be generalized. If $-\Delta+v_0(x)$ has $m$ bound states, then forming a double well $V_L$, with $L$ sufficiently large,   $H_L=-\Delta+V_L$ will have $m-$ pairs of eigenvalues: $(\Omega_{2j},\Omega_{2j+1}),\ \ j=0,\dots,m-1$,
eigenfunctions $\psi_{2j}$ (symmetric) and $\psi_{2j+1}$ anti-symmetric.
\end{example}
\bigskip
\bigskip

By Theorem \ref{theo:RW-theorem}, for small $\cN$, there exists a unique non-trivial nonlinear bound state, bifurcating from the zero solution at the ground state energy, $\Omega_0$, of $H$.  By uniqueness, ensured by the implicit function theorem, these small amplitude nonlinear bound  states have the same symmetries as the double well; they are bi-modal. We also know from \cite{AFGST:02} that for sufficiently large $\cN$ the ground state has broken symmetry. We now seek to elucidate the transition from the regime of $\cN$ small to $\cN$ large.

\bigskip

We work in the general setting of hypotheses
 {\bf (H1)-(H4)}.
  Define spectral projections onto the bound and continuous spectral
parts of $H$:
\begin{equation}
\Pzero = \left(\vzero,\cdot\right)\vzero,\ \Pone =
\left(\vone,\cdot\right)\vone,\ \ \Pc = I\ -\ \Pzero\ -\ \Pone
\label{specproj}
\end{equation}
Here,
\begin{equation}
\left(f,g\right)\ =\ \int\ \bar{f}g\ dx.\label{ip-def}
\end{equation}
We decompose the solutions of Eq. (\ref{bs-eqn}) according to
\begin{equation}
\Psi_\Omega\ =\ c_0\vzero\ +\ c_1\vone\ +\ \eta,\ \ \eta=\Pc\eta.
\label{decomp}
\end{equation}

We next substitute the expression (\ref{decomp}) into equation
(\ref{bs-eqn}) and then act with projections $\Pzero$, $\Pone$ and
$\Pc$ to the resulting equation.
Using the symmetry and anti-symmetry properties of the eigenstates, we obtain three equations which are
equivalent to the PDE (\ref{bs-eqn}):
\begin{align}
&\left(\Omega_0-\Omega\right)c_0+ a_{0000}|c_0|^2c_0 +
\left(a_{0110}+a_{0011}\right)|c_1|^2c_0 + a_{0011}c_1^2\bar c_0
+\left(\vzero g,\cR(c_0,c_1,\eta)\right)\ =\ 0\label{Pzero-eqn}\\
& \left(\Omega_1-\Omega\right)c_1+ a_{1111}|c_1|^2c_1 +
\left(a_{1010}+a_{1001}\right)|c_0|^2c_1 + a_{1010} c_0^2\bar c_1
+\left(\vone g,\cR(c_0,c_1,\eta)\right)\ =\ 0\label{Pone-eqn}\\
&\left( H - \E\right)\ \eta\ =\ -\Pc\ g\left[\ F(\cdot;c_0,c_1)\ +\
\cR(c_0,c_1,\eta)\ \right] \label{Pc-eqn}
\end{align}
$F(\cdot,c_0,c_1)$ is independent of $\eta$ and $\cR(c_0,c_1,\eta)$
contains linear, quadratic and cubic terms in $\eta$. The
coefficients
 $a_{klmn}$ are defined by:
 \begin{equation}
 a_{klmn}\ =\ \left(\ \psi_k, gK[\psi_l\psi_m]\psi_n\ \right)
 \label{aklmn}
 \end{equation}

We shall study the character of the set of solutions  of the system
(\ref{Pzero-eqn}), (\ref{Pone-eqn}), (\ref{Pc-eqn}) restricted to
the level set
\begin{equation}
\int |\Psi_\Omega|^2 dx\ =\ \cN\ \ \iff\ \ |c_0|^2\ +\ |c_1|^2\ +\ \int
|\eta|^2 dx\ =\ \cN \label{levelN}
\end{equation}
as $\cN$ varies.

Let $\Omega_0$ and $\Omega_1$ denote the two lowest eigenvalues of $H_L$. We prove (Theorem \ref{theo:symbrkbif}, Corollary \ref{cor:dwell-sb}, Theorem \ref{theo:exchange-stab}):

\begin{itemize}
\item There
exist two solution branches, parametrized by
$\cN$, which  bifurcate from the zero solution at the
eigenvalues, $\Omega_0$ and $\Omega_1$.
 \item  Along the  branch, $(\Omega,\Psi_\Omega)$,  emanating from the solution $(\Omega=\Omega_0,\Psi=0)$ , there is a symmetry breaking bifurcation at $\cN=\cN_{crit}>0$. In
particular, let $u_{crit}$ denote the solution of (\ref{bs-eqn})
corresponding to the value $\cN=\cN_{crit}$. Then, in a neighborhood
$u_{crit}$, for $\cN<\cN_{crit}$ there is only one solution of
(\ref{bs-eqn}), the symmetric ground state, while for
$\cN>\cN_{crit}$ there are two solutions one symmetric and a second
asymmetric.
\item {\it Exchange of stability} at the bifurcation
point:\ For $\cN<\cN_{crit}$ the symmetric state is dynamically
stable, while for $\cN>\cN_{crit}$ the asymmetric state is stable
and the symmetric state is exponentially unstable. \end{itemize} \bigskip

\section{ Bifurcations in a finite dimensional approximation}
\label{sec:finite-dim-approx}

It is illustrative to consider  the finite dimensional approximation to the
system (\ref{Pzero-eqn},\ref{Pone-eqn},\ref{Pc-eqn}), obtained by neglecting the
continuous spectral part, $\Pc u$. Let's first set $\eta=0$, and
therefore $\cR(c_0,c_1,0)=0$. Under this assumption of no coupling
to the continuous spectral part of $H$, we obtain the finite
dimensional system:
\begin{align}
&\left(\Omega_0-\Omega\right)c_0+ a_{0000}|c_0|^2c_0 +
\left(a_{0110}+a_{0011}\right)|c_1|^2c_0 + a_{0011}c_1^2\bar c_0
 =\ 0\label{t-Pzero-eqn}\\
& \left(\Omega_1-\Omega\right)c_1+ a_{1111}|c_1|^2c_1 +
\left(a_{1010}+a_{1001}\right)|c_0|^2c_1 + a_{1010} c_0^2\bar c_1
\ =\ 0\label{t-Pone-eqn}\\
&|c_0|^2\ +\ |c_1|^2\ =\ \cN\label{t-constraint}
\end{align}
Our strategy is to first analyze the bifurcation problem for this
approximate finite-dimensional system of {\it algebraic equations}.
 We then treat the
corrections, coming from coupling to the continuous spectral
 part of $H$,  $\eta$, perturbatively.

For simplicity we take $c_j$ real: $c_j=\rho_j\in\R$; see section
\ref{sec:bifsymbrk}. Then,
 \begin{align}
&\rho_0\ \left[\ \Omega_0-\Omega\ +\ a_{0000}\rho_0^2\ +\
\left(a_{0110}+2a_{0011}\right)\rho_1^2\ \right]  =\ 0\label{t2-Pzero-eqn}\\
& \rho_1\ \left[\ \Omega_1-\Omega\ +\ a_{1111}\rho_1^2\ +\
\left(a_{1001}+2a_{1010}\right)\rho_0^2\ \right]\ =\ 0
\label{t2-Pone-eqn}\\
&\rho_0^2\ +\ \rho_1^2\ -\ \cN\ =\ 0.\label{t2-constraint}
\end{align}
Introduce the notation
\begin{equation}
\cP_0=\rho_0^2,\ \ \cP_1=\rho_1^2 \label{P0P1}
\end{equation}
Then,
\begin{align}
&\cF_0(\cP_0,\cP_1,\E;\cN) \ =\ \cP_0\ \left[\ \Omega_0-\Omega\ +\
a_{0000}\cP_0\ +\ \left(a_{0110}+2a_{0011}\right)\cP_1\ \right]  =\
0
\nn\\
& \cF_1(\cP_0,\cP_1.\E;\cN)\ =\ \cP_1\ \left[\ \Omega_1-\Omega\ +\
a_{1111}\cP_1\ +\ \left(a_{1001}+2a_{1010}\right)\cP_0\ \right]\ =\
0
\nn\\
&\cF_{\cN}(\cP_0,\cP_1,\E;\cN)\ =\cP_0\ +\ \cP_1\ -\ \cN\ =\ 0.
\label{t3-eqns}
\end{align}

\nit{\bf Solutions of the approximate system}
\begin{itemize}
\item[(1)]\ $\cQ^{(0)}(\cN)\ =\ (\ \cP_0^{(0)},\cP_1^{(0)},\E^{(0)}\ )\ =\ (\ \cN,0,\Ezero+a_{0000}\cN\ )$\ -\ approximate nonlinear ground state branch
\item[(2)]\ $\cQ^{(1)}(\cN)\ =\ (\ \cP_0^{(1)},\cP_1^{(1)},\E^{(1)}\ )\ =\ (\ 0,\cN,\Eone+a_{1111}\cN\ )$\ -\ approximate nonlinear excited state branch
\end{itemize}

Thus we have a system of equations $\cF(\cQ,\cN)=0$, where
$\cF:(\R_+\times\R_+\times \R) \times\R_+\to\R^3\times \R_+$,
mapping $(\cQ,\cN)\to\cF(\cQ,\cN)$ smoothly. We have that
$\cF(\cQ^{(j)}(\cN),\cN)=0,\ j=0,1$ for all $\cN\ge0$. A bifurcation
(onset of multiple solutions) can occur only at a value of $\cN_*$
for which the Jacobian $d_\cQ\cF(\cQ^{(j)}(\cN_*);\cN_*)$ is
singular. The point $(\cQ^{(j)}(\cN_*);\cN_*)$ is called a {\it
bifurcation point}. In a neighborhood of a bifurcation point there
is a multiplicity of solutions (non-uniqueness) for a given $\cN$.
The detailed character of the bifurcation is suggested by the nature
of the null space of $d_\cQ\cF(\cQ^{(j)}(\cN_*);\cN_*)$.

 We next compute $d_\cQ\cF(\cQ^{(j)}(\cN);\cN)$ along the different branches in order to see whether and where there are bifurcations.
 \bigskip

The Jacobian is given by
 {{\small
\begin{align}
&d_\cQ\cF(\cQ^{(j)}(\cN);\cN)\ =\ \frac{\partial(\cF_0,\cF_1,\cF_\cN)}{\partial(\cP_0,\cP_1,\E)}\ =\nn\\
&\begin{pmatrix} \Ezero-\E+2a_{0000}\cP_0+(a_{0110}+a_{0011})\cP_1 &
(a_{0110}+2a_{0011}) \cP_0 & -P_0\\
(a_{1001}+2a_{1010})  \cP_1& \Eone-\E+ 2a_{1111}\cP_1+
(a_{0110}+2a_{1010}) \cP_0 & -\cP_1\\
1 & 1 & 0\end{pmatrix}
\label{jacobian-finite-dim}
\end{align} }}

A candidate value of $\cN$ for which there is a  bifurcation point along the ``ground state branch''  is one for which
\begin{equation}
\det\left(\ d_\cQ\cF(\cQ^{(0)}(\cN);\cN)\ \right)\ =\ 0\ \ \iff\ \
\cN\ =\
 \Ncrz\  \equiv\ \frac{\Eone-\Ezero}{a_{0000}-(a_{1001}+2a_{1010})}\label{Ncrit}
 \end{equation}
\medskip

Since the parameter $\cN$ is positive, we have
\begin{prop}\label{prop:approx-bif}
\begin{itemize}
\item[(a)]\ $\cQ^{(0)}(\Ncrz)\ =\
(\Ncrz,0,\Ezero+a_{0000}\Ncrz;\Ncrz)$ is a bifurcation point for the approximating system (\ref{t2-Pzero-eqn}-\ref{t2-constraint})  if $\Ncrz$ is positive.
\item[(b)]\ For the double well with well-separation parameter, $L$, we have that  $\Ncrz(L)>0$ for $L$ sufficiently large.
\end{itemize}
\end{prop}
\medskip

\nit{\bf Proof:}\ We need only check (b). This is easy to see, using
 the large $L$ approximations of $\psi_0$ and $\psi_1$ in terms of $\psi_\omega$, the ground state of $H=-\Delta+V(x)$, the ``single well''
 operator:
 \begin{align}
\vzero\ &\sim\ 2^{-1/2}\left(\ \psi_\omega(x-L)+\psi_\omega(x+L)\
\right)\nn\\\vone\ &\sim\ 2^{-1/2}\left(\
\psi_\omega(x-L)-\psi_\omega(x+L)\ \right);
\label{psilargeL}\end{align}
see Proposition \ref{dw1} in section
\ref{sec:doublewells}.
\bigskip

\nit{\bf Excited state branch}
\begin{equation}
\det\left(\ d_\cQ\cF(\cQ^{(1)}(\cN);\cN)\ \right)\ =\ 0\ \iff\ \
 \cN_*^{(1)}\ =\ \frac{\Eone-\Ezero}{a_{0110}+2a_{0011}-a_{1111}}
 \label{nobif}
\end{equation}
\begin{rmk}  For the double well with well-separation parameter, $L$, we have that  $\cN^{(1)}_*(L)~<~0$ for $L$ sufficiently large, as can be checked using the approximation (\ref{psilargeL}). Therefore $\cQ^{(1)}(\cN_*^{(1)})$ is {\bf not} a bifurcation point of
the approximating system   (\ref{t3-eqns}).
\end{rmk}
\medskip

\nit{\bf Summary:}\ Assume $\cN$ is sufficiently small.
 The finite dimensional approximation (\ref{t3-eqns}) predicts a
 symmetry breaking bifurcation along the  nonlinear ground state
 branch and that no bifurcation takes place along the anti-symmetric
 branch of nonlinear bound states.

 \section{Bifurcation / Symmetry breaking analysis of the PDE}
 \label{sec:bifsymbrk}

In this section we prove the following
 \begin{theo}\label{theo:symbrkbif} {\bf (Symmetry Breaking for NLS-GP)}\
 Consider NLS-GP with hypotheses {\bf (H2)-(H4)}.  Let $a_{klmn}$ be given by (\ref{aklmn}) and
\begin{align}
 \Xi[\psi_0,\psi_1,g]\ &\equiv a_{0000}-a_{1001}-2a_{1010}\nn\\
  &=
    \left(\psi_0^2,gK[\psi_0^2]\right)- \left(\psi_1^2,gK[\psi_0^2]\right)-2 \left(\psi_0\psi_1,gK[\psi_0\psi_1]\right)\ >\ 0.
  \label{Xi-def}\end{align}
  Assume
 \begin{equation}
  \frac{\Omega_1-\Omega_0}{\Xi[\psi_0,\psi_1]^2}\ \
 {\rm is\ sufficiently\ small.} \label{hyp:closeev}\end{equation}
 Then, there exists $\cN_{cr}>0$ such that
 \begin{itemize}
 \item[(i)]\ for any $\cN\le\cN_{cr}$, there is (up to the symmetry $u\mapsto u\ e^{i\gamma}$) a unique ground state, $u_\cN$, having the same spatial symmetries as the double well.
 \item[(ii)]\ $\cN=\cN_{cr},\ u^{sym}_{\cN_{cr}}$ is a bifurcation point.
 For $\cN>\cN_{cr}$, there are, in a neighborhood of
 $\cN=\cN_{cr},\ u^{sym}_{\cN_{cr}}$, two branches of solutions: (a) a continuation of the symmetric branch, and (b) a new asymmetric branch.
 \item[(iii)]\ The critical $\cN$- value for bifurcation is given approximately by
\begin{align}
\cN_{cr}\ &\ =\ \frac{\Omega_1-\Omega_0}{\Xi[\psi_0,\psi_1]}
 \left[1+ \cO\left(\
  \frac{\Omega_1-\Omega_0}{\Xi[\psi_0,\psi_1]^{2}}\right)\right]
  \nn\end{align}
   \end{itemize}
 \end{theo}
 \begin{cor}\label{cor:dwell-sb}
Fix a pair of eigenvalues, $(\Omega_{2j},\psi_{2j}),\
 (\Omega_{2j+1},\psi_{2j+1})$ of the linear double-well potential, $V_L(x)$; see 
  Example \ref{example:basic-example}.
 For the NLS-GP with double well potential of well-separation $L$,
 there exists $\tilde L>0$, such that for all $L\ge\tilde L$, there is a symmetry breaking bifurcation, as described in Theorem \ref{theo:symbrkbif}, with  $\cN_{cr}=\cN_{cr}(L;j)$.
 \end{cor}\label{rem:Llarge}
 \begin{rmk} $\Omega_1(L)-\Omega_0(L)\ =\ \cO(e^{-\kappa L})$ for $L$ large. The terms in  $\Xi[\psi_0,\psi_1](L)$ are $\cO(1)$. Therefore,
  for the double well potential, $V_L(x)$, the smallness hypothesis of Theorem \ref{theo:symbrkbif}  holds provided $L$ is sufficiently large.
 \end{rmk}

To prove this theorem we will establish that, under
hypotheses
\eqref{Xi-def}-\eqref{hyp:closeev}, the character of the solution
set (symmetry breaking bifurcation) of the finite dimensional
approximation
 (\ref{t-Pzero-eqn}-\ref{t-constraint}) persists for the full (infinite dimensional) problem:
  \begin{align}
&\left(\Omega_0-\Omega\right)c_0+ a_{0000}|c_0|^2c_0 +
\left(a_{0110}+a_{0011}\right)|c_1|^2c_0 + a_{0011}c_1^2\bar c_0
+\left(\vzero g,\cR(c_0,c_1,\eta)\right)\ =\ 0\label{Pzero-eqn1}\\
& \left(\Omega_1-\Omega\right)c_1+ a_{1111}|c_1|^2c_1 +
\left(a_{1010}+a_{1001}\right)|c_0|^2c_1 + a_{1010} c_0^2\bar c_1
+\left(\vone g,\cR(c_0,c_1,\eta)\right)\ =\ 0\label{Pone-eqn1}\\
&\left( H - \E\right)\ \eta\ =\ -\Pc\ g\left[\ F(\cdot;c_0,c_1)\ +\ \cR(c_0,c_1,\eta)\ \right],\ \ \ \ \ \eta=\Pc\eta \label{Pc-eqn1}\\
&|c_0|^2\ +\ |c_1|^2\ +\ \int |\eta|^2\ \ =\ \cN. \label{constraint}
\end{align}

We analyze this system using the Lyapunov-Schmidt method. The
strategy is to solve equation  (\ref{Pc-eqn1}) for $\eta$ as a
functional of $c_0,c_1$ and $\Omega$. Then, substituting
$\eta=\eta[c_0,c_1,\Omega]$ into equations (\ref{Pzero-eqn1}),
(\ref{Pone-eqn1}) and (\ref{constraint}), we obtain three closed
equations, depending on a parameter $\cN$, for $c_0,c_1$ and
$\Omega$. This system is a perturbation of the  finite dimensional
(truncated) system:\ (\ref{t-Pzero-eqn}, \ref{t-Pone-eqn}) and
(\ref{t-constraint}). We then show that under hypotheses
\eqref{Xi-def}-\eqref{hyp:closeev} there is a symmetry breaking
bifurcation. Finally, we show that the terms perturbing the finite
dimensional model have a small and controllable effect on the
character of the solution set for a range of $\cN$, which includes
the bifurcation point. Note that, in the double well problem,
hypotheses \eqref{Xi-def}-\eqref{hyp:closeev} are satisfied for $L$
sufficiently large, see Proposition \ref{dw2}.
\bigskip

We begin with the following proposition, which characterizes
$\eta=\eta[c_0,c_1,\Omega]$.
\begin{prop}\label{prop:eta}
Consider equation (\ref{Pc-eqn1}) for $\eta$.  By {\bf (H4)} we have the following:
\begin{equation}
 {\rm Gap\ Condition:}\ |\Omega_j-\tau|\ \ge\ 2d_*\ {\rm  for}\ j=0,1\ {\rm and\ all}\  \
   \tau\in \sigma(H)\setminus\{\Omega_0,\Omega_1\}
   \label{gapcond}
   \end{equation}

\nit Then there exists $n_*,\ r_*>0$, depending on $d_*$,  such that in the open set
 \ba
 |c_0|+|c_1|&<&r_*\label{c0c1r}\\
 \|c_0\vzero+c_1\vone+\eta\|_{H^2}&<&n_*(d_*)\nn\\
 dist(\Omega,\sigma(H)\setminus\{\Omega_0,\Omega_1\})&>&d_*,
 \label{Omegad}
 \ea
the unique solution of \eqref{Pc-eqn} is given by the real-analytic
mapping:
 \be (c_0,c_1,\Omega)\ \mapsto\ \ \eta[c_0,c_1,\Omega],
  \label{eta01omega}\ee
  defined on the domain given by (\ref{c0c1r},\ref{Omegad}).
Moreover there exists $C_*>0$ such that:
\begin{equation}\label{eta-est}
  \|\  \eta[c_0,c_1,\Omega]\ \|_{H^2}\le C_*(|c_0|+|c_1|)^3
\end{equation}
 \end{prop}
 \nit{\bf Proof:}\ Consider the map
 \[ N:H^2\times H^2\times H^2\mapsto L^2\]
 \[ N(\phi_0,\phi_1,\phi_2)=gK[\phi_1\phi_2]\phi_3.\]
By assumptions on the nonlinearity (see section \ref{sec:tformulation}),
there exists a
 constant $k>0$ such that
 \be\label{N-est} \|N(\phi_0,\phi_1,\phi_2)\|_{L^2}\le
 k\|\phi_1\|_{H^2}\|\phi_2\|_{H^2}\|\phi_3\|_{H^2}.
 \ee
 Moreover the map being linear in each component it is real analytic.
 \footnote{The
 trilinearity follows from the implicit bilinearity of $K$ in
 formulas \eqref{Pzero-eqn}-\eqref{Pc-eqn}.}

Let $c_0,c_1$ and $\Omega$ be restricted according the
inequalities (\ref{c0c1r},\ref{Omegad}).
 Equation (\ref{Pc-eqn}) can be rewritten in the form
 \begin{equation}\label{eta-eqn}
 \eta+(H-\Omega)^{-1}\Pc
 N[c_0\vzero+c_1\vone+\eta]\ =\ 0.
 \end{equation}
Since the spectrum of $H\Pc $ is  bounded away from $\Omega$ by $d_*$, the
resolvent:
\[(H-\Omega)^{-1}\Pc:L^2\mapsto H^2\]
is a (complex) analytic map and   bounded uniformly,
\begin{equation}\label{res-ubd}
\|(H-\Omega)^{-1}\Pc\|_{L^2\mapsto H^2}\le p(d_*^{-1}),
\end{equation}
where $p(s)\to\infty$ as $s\to\infty$.
Consequently the map $F:\mathbb{C}^2\times\{\Omega\in\mathbb{C}\ :\
dist(\Omega,\sigma(H)\setminus\{\Omega_0,\Omega_1\}\ \}\ge d_*\}\times H^2\mapsto H^2$ given by
\begin{equation}
F(c_0,c_1,\Omega,\eta)=\eta+(H-\Omega)^{-1}\Pc
 N[\ c_0\vzero+c_1\vone+\eta\ ]
 \label{Fdef}
 \end{equation}
 is real analytic. Moreover,
 \begin{equation}
 F(0,0,\Omega,0)=0,\ \ \
 D_\eta F(0,0,\Omega,0)=I.\nn
 \end{equation}
Applying the implicit function theorem to equation \eqref{eta-eqn},
we have that  there exists $n_*(\Omega),\ r_*(\Omega)$ such that whenever $|c_0|+|c_1|<r_*$ and
$\|c_0\vzero+c_1\vone+\eta\|_{H^2}<n_*$ equation \eqref{eta-eqn} has
an unique solution:
\[\eta=\eta(c_0,c_1,\Omega)\in H^2\] which depends analytically on the parameters
$c_0,\ c_1,\ \Omega.$ By applying the projection operator $\Pc $ to
the \eqref{eta-eqn} which commutes with $(H-\Omega)^{-1}$ we
immediately obtain $\Pc\eta=\eta,$ i.e. $\eta\in \Pc L^2.$

We now show that $n_*,\ r_*$ can be chosen independent of
$\Omega$, satisfying  (\ref{Omegad}). The implicit function theorem can be
applied in an open set for which
\begin{equation}
D_\eta F(c_0,c_1,\Omega,\eta)=I+(H-\Omega)^{-1}\Pc D_\eta
N[\ c_0\vzero+c_1\vone+\eta\ ]
\nn
\end{equation}
is invertible. For this it suffices to have:
 \[\|\ (H-\Omega)^{-1}\Pc D_\eta
 N[\ c_0\vzero+c_1\vone+\eta]\ \|_{H^2}\le
 Lip<1\]
 A direct application of \eqref{N-est} and \eqref{res-ubd} shows that
\[
\|(H-\Omega)^{-1}\Pc D_\eta
 N[\ c_0\vzero+c_1\vone+\eta]\ \|_{H^2}\le\]
\begin{equation}
 3k\ p(d_*^{-1})\ \|c_0\vzero+c_1\vone+\eta\|_{H^2}^2\label{lip-est}
\end{equation}

Fix
 $Lip=3/4.$ Then,
a sufficient condition for invertibility is
\begin{equation}\label{r-est}
3k\ p(d_*^{-1})\ \|c_0\vzero+c_1\vone+\eta\|_{H^2}^2\le Lip=3/4.
\end{equation} which allows us to choose
  $n_*=\frac{1}{2}\sqrt{\frac{1}{k p(d_*^{-1})}}$,
independently of $\Omega.$

 But, if \eqref{r-est} holds, then, from
\eqref{lip-est}, the $H^2$ operator
\begin{equation}
 (H-\Omega)^{-1}\Pc
 N[\ c_0\vzero+c_1\vone+\cdot\ ]
 \nn\end{equation}
 is Lipschitz with Lipschitz constant less or equal to $Lip=3/4.$ The standard contraction principle estimate applied
 to \eqref{eta-eqn} gives:
 \begin{eqnarray}
 \|\eta\|_{H^2}&\le &\frac{1}{1-Lip}\|(H-\Omega)^{-1}\Pc
 N[\ c_0\vzero+c_1\vone\ ]\ \|_{H^2}\nn\\
 &\le &
 4 p(d_*^{-1})\ k \|c_0\vzero+c_1\vone\|_{H^2}^3.\label{eta-est1}
 \end{eqnarray}

Plugging the above estimate into \eqref{r-est} gives:
\[\|c_0\vzero+c_1\vone\|_{H^2}+4p(d_*^{-1})\ k\|c_0\vzero+c_1\vone\|_{H^2}^3\le
\frac{1}{2\sqrt{p(d_*^{-1}) k}}\] Since the left hand side is continuous in
$(c_0,c_1)\in\mathbb{C}^2$ and zero for $c_0=c_1=0$ one can
construct $r_*>0$ depending only on $d_*,\ k$ such that the above
inequality, hence \eqref{r-est} and \eqref{eta-est1}, all hold
whenever $|c_0|+|c_1|\le r_*.$
 Finally, \eqref{eta-est} now follows from \eqref{eta-est1}.QED

  \medskip

In particular, for the double well potential we have the following
  \begin{prop} Let $V=V_L$ denote the double well potential with well-separation
  $L$. There exists $ L_*>0,\ \veps(L_*)>0$ and $d_*(L_*)>0$ such that for $L>L_*,$
  we have that for $(c_0,c_1,\Omega)$ satisfying
  $dist(\Omega,\sigma(H)\setminus\{\Omega_0,\Omega_1\}\}\ \}\ge d_*(L_*)$ and $|c_0|+|c_1|<\veps(L_*)$
  $\eta[c_0,c_1,\Omega]$ is defined  and analytic and satisfies the bound (\ref{eta-est}) for some $C_*>0$.
  \end{prop}

  \nit{\bf Proof}:
  Since $\Omega_0,\ \Omega_1,\ \vzero$ and $\vone$ can be controlled, uniformly in $L$ large,
  via the approximations (\ref{psilargeL}), both $d_*$ and $r_*$ in the previous
  Proposition can be controlled uniformly in $L$ large. QED
  \medskip

Next we study the symmetries of $\eta(c_0,c_1,\Omega)$ and
properties of $\cR(c_0,c_1,\eta)$ which we will use in analyzing the
equations \eqref{Pzero-eqn}-\eqref{Pone-eqn}. The following result
is a direct consequence of the symmetries of equation \eqref{Pc-eqn}
and Proposition \ref{prop:eta}:
\begin{prop}\label{prop:sym}
We have
\begin{align}
&\eta(e^{i\theta}c_0,e^{i\theta}c_1,\Omega)=e^{i\theta}\eta(c_0,c_1,\Omega),
\qquad {\rm for}\  0\le\theta<2\pi ,\label{rot}\\
&\overline{\eta(c_0,c_1,\Omega)}=\eta(\overline{c_0},\overline{c_1},\overline\Omega)\label{cc}
\end{align}
in particular
\begin{align}
&\eta(e^{i\theta}c_0,c_1=0,\Omega)=e^{i\theta}\eta(c_0,c_1=0,\Omega),\label{rot0}\\
&\eta(c_0=0,e^{i\theta}c_1,\Omega)=e^{i\theta}\eta(c_0=0,c_1,\Omega),\label{rot1}
\end{align}
$\eta(c_0,0,\Omega)$ is even in $x_1,$ $\eta(0,c_1,\Omega)$ is odd
in $x_1$ and if $c_0,\ c_1$ and $\Omega$ are real valued, then
$\eta(c_0,c_1,\Omega)$ is real valued.

In addition
\begin{align}
&\langle\ \vzero , \cR(c_0, c_1,\eta)\ \rangle\ =\ c_0\
 f_0\left(\ c_0,c_1,\Omega\ \right)\label{f0}\\
& \langle\ \vone , \cR(c_0, c_1,\eta)\ \rangle\ =\ c_1\ f_1\left(\
c_0,c_1,\Omega\ \right) \label{f1}
\end{align}
where, for any $0\le\theta<2\pi$
 \begin{align}
 &f_j\left(e^{i\theta}c_0,e^{i\theta}c_1,\Omega\ \right)=f_j\left(\ c_0,c_1,\Omega\ \right), \qquad j=0,1 \label{f0f1-rot}\\
 &\overline{f_j(c_0,c_1,\Omega\ )}=f_j\left(\ \overline{c_0},\overline{c_1},\overline\Omega\ \right), \qquad j=0,1 \label{f0f1-cc}\\
 & \left| f_j\left(\ c_0,c_1,\Omega\ \right)\right|\le
 C(|c_0|+|c_1|)^4,\qquad j=0,1 \label{f0f1-est}
\end{align}
for some constant $C>0.$ Moreover, both $f_0$ and $f_1$ can be
written as absolutely convergent power series:
 \begin{equation}\label{f0f1-ser}
 f_j(c_0,c_1,\Omega)=\sum_{k+l+m+n\ge 4,\ k-l+m-n=0,\ m+n={\rm
 even}}b^j_{klmn}(\Omega)c_0^k\ \overline c_0^l\ c_1^m\ \overline c_1^n, \qquad j=0,1,
 \end{equation}
where $b^j_{klmn}(\Omega)$ are real valued when $\Omega$ is real
valued. In particular, if $c_0,\ c_1$ and $\Omega$ are real valued,
then $f_j\left(c_0,c_1,\Omega\ \right)$ is real valued and, in polar
coordinates, for $c_0,c_1\not= 0,$ we have
\begin{equation}\label{f0f1-ser1}
f_j(|c_0|,|c_1|,\dph ,\Omega)=\sum_{k+m\ge 2,\
p\in\mathbb{Z}}b^j_{kmp}(\Omega)e^{ip2\dph}|c_0|^{2k}|c_1|^{2m},
\qquad j=0,1,
\end{equation}
where $\dph$ is the phase difference between $c_1\in\mathbb{C}$ and
$c_0\in\mathbb{C}.$
\end{prop}

\nit{\bf Proof of Proposition \ref{prop:sym}:}\ We start with
\eqref{rot} which clearly implies \eqref{rot0}-\eqref{rot1}. We fix
$\Omega$ and suppress dependence on it in subsequent notation.
From equation \eqref{eta-eqn} we have:
 \ba
&&\eta(e^{i\theta}c_0,e^{i\theta}c_1)\nn\\
&=&-(H-\Omega)^{-1}\Pc
N(e^{i\theta}c_0\vzero+e^{i\theta}c_1\vone+\eta,e^{i\theta}c_0\vzero+e^{i\theta}c_1\vone+\eta,e^{i\theta}c_0\vzero+e^{i\theta}c_1\vone+\eta)\nn\\
&=&-(H-\Omega)^{-1}\Pc e^{i\theta}
N(c_0\vzero+c_1\vone+e^{-i\theta}\eta,c_0\vzero+c_1\vone+e^{-i\theta}\eta,c_0\vzero+c_1\vone+e^{-i\theta}\eta)\nn
\ea where we used \be\label{N-gauge}
N(a\phi_1,b\phi_2,c\phi_3)=a\overline b cN(\phi_1,\phi_2,\phi_3).\ee
Consequently
\[e^{-i\theta}\eta(e^{i\theta}c_0,e^{i\theta}c_1)=-(H-\Omega)^{-1}\Pc
N[\ c_0\vzero+c_1\vone+e^{-i\theta}\eta,c_0\vzero+c_1\vone+e^{-i\theta}\eta,c_0\vzero+c_1\vone+e^{-i\theta}\eta]
\]
which shows that both
$e^{-i\theta}\eta(e^{i\theta}c_0,e^{i\theta}c_1)$ and
$\eta(c_0,c_1)$ satisfy the same equation \eqref{eta-eqn}. From the
uniqueness of the solution proved in Proposition \ref{prop:eta} we
have the relation \eqref{rot}.

A similar argument
(and use of the complex conjugate)
leads to \eqref{cc} and to the parities of
$\eta(c_0,0)$ and $\eta(0,c_1).$

To prove \eqref{f0} and \eqref{f1}, recall that \ba
\cR\left(c_0,c_1,\eta(c_0,c_1,\Omega)\right)&=&N(c_0\vzero+c_1\vone+\eta,c_0\vzero+c_1\vone+\eta,c_0\vzero+c_1\vone+\eta)\nn\\
&-&N(c_0\vzero+c_1\vone,c_0\vzero+c_1\vone,c_0\vzero+c_1\vone).\label{cr-def}
\ea Consider first the case $c_1=\rho_1\in\mathbb{R}.$ Note that
\[ \langle\ \vone g, \cR(c_0,
\rho_1=0,\eta(c_0,0))\ \rangle\ =\ 0.\] Indeed, for $\rho_1=0,$ all
the functions in the arguments of $\cR $ are even functions (in
$x_1$) making $\cR$ an even function. Since $\vone$ is odd we get
that the above is the integral over the entire space of an odd
function, i.e. zero. Since $\langle\ \vone , \cR(c_0,
\rho_1,\eta(c_0,\rho_1))\ \rangle$ is analytic in
$\rho_1\in\mathbb{R}$ by the composition rule, and its Taylor series
starts with zero we get \eqref{f1} for real $c_1=\rho_1.$ To extend
the result for complex values $c_1$ we use the rotational symmetry
 of $\cR,$ namely from \eqref{rot},
\eqref{N-gauge} and \eqref{cr-def} we have
\[\cR\left(e^{i\theta}c_0,e^{i\theta}c_1,\eta(e^{i\theta}c_0,e^{i\theta}c_1,\Omega)\right)=e^{i\theta}\cR\left(c_0,c_1,\eta(c_0,c_1,\Omega)\right),
\qquad 0\le\theta<\pi\] hence \eqref{f1} holds for
$c_1=|c_1|e^{-i\theta}$ by extending $f_1$ via the equality
\eqref{f0f1-rot}.

A very similar argument holds for \eqref{f0}. Equation
\eqref{f0f1-est} follows from the definition of $\cR$ and
\eqref{eta-est}. Equation \eqref{f0f1-cc} follows from \eqref{cc}.

We now turn to a proof of the expansions for $f_j$: (\ref{f0f1-ser})
and (\ref{f0f1-ser1}).
 Note first that both $f_0$ and $f_1$ are real analytic in $c_0,\ c_1$  by analyticity of
$\cR$ in \eqref{f0}-\eqref{f1}; see \eqref{cr-def}.
 Note also that
$\eta$ is real analytic by Proposition \ref{prop:eta} while $N$ is
trilinear. Hence, both $f_0$ and $f_1$ can be written in power
series of the type \eqref{f0f1-ser}. Estimate \eqref{f0f1-est}
implies that $k+l+m+n\ge 4,$ while the rotational invariance
\eqref{f0f1-rot} implies $k-l+m-n=0.$ The following parity argument
shows why $m+n$ hence $m-n=l-k$ and $k+l$ are all even. Assume $m+n$
is odd. Note that because of \eqref{f0}, $b^0_{klmn}$ is the scalar
product between an even function (in $x_1$) $\vzero$ and the term in
the power series of $\cR$ in which $\vone$ is repeated $m+n$ times.
The latter is an odd function (in $x_1$) because $\vone$ is an odd
function and it is repeated an odd number of times. The scalar
product and hence $b^0_{klmn}$ for $m+n$ odd will be zero. A similar
argument holds for $b^1_{klmn},\ m+n$ odd. Finally
$b^j_{klmn}(\Omega)$ are real valued when $\Omega$ is real because
they are scalar products of real valued functions.

The form \eqref{f0f1-ser1} of the power series follows directly from
\eqref{f0f1-ser} by expressing $c_0$ and $c_1$ in their polar forms:
$c_0=|c_0|e^{i\theta_0}$ and $c_1=|c_1|e^{i\theta_1},\
\dph=\theta_1-\theta_0$, and using that $m+n,\ k+l$ and $m-n=-(k-l)$
are all even.
  The proof of Proposition \ref{prop:sym} is now complete.

\subsection{Ground state and excited state branches, pre-bifurcation }

In this section we prove part (i) of Theorem \ref{theo:symbrkbif}
 as well as a corresponding statement about the excited state. In particular, we show that for sufficiently small amplitude, the only nonlinear bound state families are those arising via bifurcation from the zero state at the eigenvalues $\Omega_0$ and $\Omega_1$. This
 is true for general potentials with two bound states. Here, however we can determine threshold amplitude, $\cN_{cr}$, above which the solution set changes.

A closed system of equations for $c_0,c_1$ and $\Omega$,
parametrized by $\cN$, is obtained upon substitution of $\eta[c_0,c_1,\Omega]$,
(Proposition \ref{prop:eta}) into
(\ref{Pzero-eqn1}-\ref{constraint}).  Furthermore, using the
structural properties  (\ref{f0}-\ref{f1}) of Proposition
\ref{prop:sym},
 we obtain:
\begin{align}
&\left(\Omega_0-\Omega\right)c_0+ a_{0000}|c_0|^2c_0 +
\left(a_{0110}+a_{0011}\right)|c_1|^2c_0 + a_{0011}c_1^2\bar c_0
+c_0f_0(c_0,c_1,\Omega)\ =\ 0\label{Pzero-eqn2}\\
& \left(\Omega_1-\Omega\right)c_1+ a_{1111}|c_1|^2c_1 +
\left(a_{1010}+a_{1001}\right)|c_0|^2c_1 + a_{1010} c_0^2\bar c_1
+c_1f_1(c_0,c_1,\Omega)\ =\ 0\label{Pone-eqn2}\\
& |c_0|^2\ +\ |c_1|^2\ + \cO\left(|c_0|^2+|c_1|^2\right)^3\ =\ \cN
\label{N-eqn2}
\end{align}
{\it This system of equations is valid for $|c_0|+|c_1|<r_*$,
independent of $L$, the distance between wells.}

If we choose $c_1=0$, then the second equation in the system
(\ref{Pzero-eqn2}) is satisfied. In this case, a non-trivial
solution requires $c_0\ne0$. The first equation, (\ref{Pone-eqn2}),
after factoring out $c_0$ becomes
 \begin{equation}
 \Omega_0-\Omega+ a_{0000}|c_0|^2+f_0(|c_0|,0,\Omega)\ =\ 0
 \label{gsb-eqn}
 \end{equation}
where we used \eqref{f0f1-rot} to eliminate the phase of the complex
quantity $c_0$. Since  $\Omega$ is real
%
%
\eqref{gsb-eqn} becomes one equation with two real parameters
$\Omega,\ |c_0|.$ Since the right hand side of \eqref{gsb-eqn}
vanishes for $\Omega=\Omega_0$ and $|c_0|=0$ and since the partial
derivative of this function with respect to $\Omega$, evaluated
at this solution, is non-zero, we have by the implicit function
theorem that there is a unique solution \be
\Omega=\Omega_g(|c_0|)=\Omega_0+a_{0000}|c_0|^2+\cO (|c_0|^4).
\label{Omega-g}\ee By (\ref{N-eqn2}), for small amplitudes, the
mapping from  $|c_0|^2+|c_1|^2$ to $\cN$ is invertible. The  family
of solutions
\begin{equation}
|c_0|\ \mapsto\ \left(\ |c_0|e^{i\theta},
|c_1|=0,\Omega=\Omega_g(|c_0|)\ \right),\ \ \theta_0\in[0,2\pi)
\nn\end{equation}
 {\it defined for $|c_0|$ sufficiently small},   corresponds to a family of symmetric nonlinear bound states, $u_\cN$
with
$\|u_\cN\|_{L^2}^2=\cN$, bifurcating  from the zero solution at the
linear eigenvalue $\Omega_0$
\begin{align} u_\cN\ &=\ \left(\
|c_0|\vzero(x)\ +\ \eta[|c_0|,0,\Omega_g(|c_0|](x)\ \right)
 \ e^{i\theta_0},\ \ \theta_0\in [0,2\pi)\nn\\
\Omega\ &=\ \Omega_g(|c_0|);\nn
\end{align}
see, for example,  \cite{pw:97, RW:88}.
 Since both $\vzero$ and
$\eta(|c_0|,0,\Omega_g)$ are even (in $x_1$) we infer that $u_\cN$
is
 symmetric (even).

\begin{rmk}\label{rmk:antisym1}
 A similar result holds for the case $c_0=0$ leading to the anti-symmetric excited state branch.
\end{rmk}

\begin{prop}\label{prop:unique} For $|c_0|+|c_1|$ sufficiently small, these two branches of solutions,  are the only solutions non-trivial solutions of (\ref{bs-eqn}).
\end{prop}

\nit {\bf Proof:}\  Indeed,  suppose the contrary.  By local uniqueness of these branches, ensured by the implicit function theorem, a solution not already lying on one of these branches must have both $c_0$ and $c_1$ nonzero. Now, divide the first equation by $c_0$, the second equation by  $c_1,$ and subtract the
results. Introducing polar coordinates:
 \begin{equation}
 c_0=\rho_0e^{i\theta_0},\ c_1=\rho_1e^{i\theta_1},\
 \dph=\theta_1-\theta_0,\label{polc0}
 \end{equation}
 we obtain:
 \begin{eqnarray}
    \Omega_1-\Omega_0&=&a_{0000}\rho_0^2\ +\
 \left(a_{0110}+a_{0011}+a_{0011}e^{i2\dph}\right)\rho_1^2\ +\ f_0\left(\
 \rho_0,\rho_1,\dph ,\Omega\right)\nn\\
 &-&a_{1111}\rho_1^2\ -\ \left(a_{1001}+a_{1010}+a_{1010}e^{-i2\dph}\right)\rho_0^2\ -\ f_1\left(\
 \rho_0,\rho_1,\dph ,\Omega\right).\label{omegarho-small}
 \end{eqnarray}
The left hand side is nonzero while the right hand side is
continuous, uniformly for $\Omega$  satisfying (\ref{Omegad}) and zero for
$\rho_0=0=\rho_1.$  Equation (\ref{omegarho-small}) cannot hold for
$|\rho_0|+|\rho_1|<\veps $ where $\veps >0$ is independent of
$\Omega.$ This completes the proof of Proposition \ref{prop:unique}.
\bigskip

Note, however that nothing can prevent \eqref{omegarho-small} to
hold for larger $\rho_0$ and $\rho_1$ possibly leading to a third
branch of solutions of
(\ref{bs-eqn}). In what follows, we show that this is
indeed the case and the third branch bifurcates from the ground
state one at a critical value of $\rho_0=\rho_0^*.$

\subsection{Symmetry breaking bifurcation along the ground state / symmetric branch}\label{sec:symbrkbifur}
A consequence of the previous section is  that there are no
bifurcations from the ground state branch for sufficiently small
amplitude. We now show seek a bifurcating branch of solutions to
(\ref{Pzero-eqn}-\ref{N-eqn2}), along which $c_0\cdot c_1\ne0$. As
argued just above, along such a new branch one must have:
 \begin{align}
 &\Omega_0-\Omega+ a_{0000}\rho_0^2 +
 \left(a_{0110}+a_{0011}+ a_{0011}e^{i2\dph}\right)\rho_1^2
 +f_0(\rho_0,\rho_1,\dph ,\Omega)\ =\ 0\label{f3-Pzero-eqn}\\
 & \Omega_1-\Omega+ a_{1111}\rho_1^2 +
 \left(a_{1010}+a_{1001}+ a_{1010}e^{-i2\dph}\right)\rho_0^2
 +f_1(\rho_0,\rho_1,\dph ,\Omega)\ =\ 0\label{f3-Pone-eqn}
 \end{align}

We first derive constraints on $\Delta\theta$. Consider the
imaginary parts of the two equations and use the expansions
\eqref{f0f1-ser1} and the fact that $\Omega$ is real:
 \begin{align}
 &a_{0011}\sin(2\dph)\rho_1^2+\sum_{k+m\ge 2,\
 p\in\mathbb{Z}}b^0_{kmp}(\Omega)\sin(p2\dph )\rho_0^{2k}\rho_1^{2m}=0\nn\\
 &a_{1010}\sin(2\dph)\rho_0^2+\sum_{k+m\ge 2,\
 p\in\mathbb{Z}}b^1_{kmp}(\Omega)\sin(p2\dph)\rho_0^{2k}\rho_1^{2m}=0.\nn
 \end{align}
Since both left hand sides are convergent series in $\rho_0,\
\rho_1,$ then all their coefficients must be zero. Hence
$\sin(2\dph)=0$ or, equivalently:
 \be\label{dphase-eqn}
 \dph\in\left\{0,\frac{\pi}{2},\pi,\frac{3\pi}{2}\right\}
 \ee

\nit{\bf Case 1:\ $\dph\in\{0,\pi\}$:}\

\nit Here, the system \eqref{f3-Pzero-eqn}-\eqref{f3-Pone-eqn} is
equivalent with the same system of two real equations with three
real parameters $\rho_0\ge 0,\ \rho_1\ge 0$ and $\Omega :$
 \begin{align}
 &F_0(\rho_0,\rho_1,\Omega)\stackrel{def}{=}\Omega_0-\Omega+ a_{0000}\rho_0^2 +
 \left(a_{0110}+2a_{0011}\right)\rho_1^2
 +f_0(\rho_0,\rho_1 ,\Omega)\ =\ 0\label{f4-Pzero-eqn}\\
 &F_1(\rho_0,\rho_1,\Omega)\stackrel{def}{=}\Omega_1-\Omega+ a_{1111}\rho_1^2 +
 \left(2a_{1010}+a_{1001}\right)\rho_0^2
 +f_1(\rho_0,\rho_1 ,\Omega)\ =\ 0\label{f4-Pone-eqn}
 \end{align}
We shall prove that there is a bifurcation point along the symmetric
branch using \eqref{Xi-def}-\eqref{hyp:closeev}, which depend on
discrete eigenvalues and eigenstates of $-\Delta+V(x).$
 These properties are proved for the double well
in section
 \ref{sec:doublewells}, an  Appendix on double wells.
\bigskip

We begin by  seeking the point along the ground state branch
$(\rho_0^*,0,\Omega_g(\rho_0^*))$ from which a new  family of
solutions of \eqref{f4-Pzero-eqn}-\eqref{f4-Pone-eqn}, parametrized
by $\rho_1\ge0$,  bifurcates; see \eqref{Omega-g}.

Recall first that for {\it any} $\rho_0\ge0$ sufficiently small,
 $F_0\left(\rho_0,0,\Omega_g(\rho_0)\right)=0$. A candidate for a bifurcation point is
$\rho_0^*>0$ for which, in addition,
 \be\label{rhocrit-eqn}
 F_1(\rho_0^*,0,\Omega_g(\rho_0^*))=0\ee
 Using \eqref{Xi-def} and \eqref{hyp:closeev} one can check that
 \be\label{rhocrit-eqn1}
 F_1(\rho_0,0,\Omega_g(\rho_0))=\Omega_1-\Omega_0+\left(a_{1001}+2a_{1010}-a_{0000}+\cO(\rho_0
 ^2)\right)\rho_0^2=0\ee
 has a solution:
 \begin{equation}\label{rhocrit:app}
 \rho_0^*\ =\ \sqrt{\frac{\Omega_1-\Omega_0}{|a_{1001}+2a_{1010}-a_{0000}|}}\ \left[1+\cO\left(\ \frac{\Omega_1-\Omega_0}{|a_{1001}+2a_{1010}-a_{0000}|^2}\
 \right)\right]
 \end{equation}
We now show that a new family of solutions bifurcates from the
symmetric state at $(\rho_0^*,0,\Omega_g(\rho_0^*))$. This is
realized as  a unique, one-parameter family of solutions
\begin{equation}
\rho_1\mapsto (\rho_0(\rho_1),\rho_1,\Omega_{asym}(\rho_1))
\label{asym-branch}
\end{equation}
 of the equations:
\begin{equation}
F_0(\rho_0,\rho_1,\Omega)=0,\ \ \ F_1(\rho_0,\rho_1,\Omega)=0
\label{F1F2eq0}
\end{equation}

 To see this, note that by the preceding discussion we have
 $F_j(\rho_0^*,0,\Omega_g(\rho_0^*))=0, j=1,2.$ Moreover, the Jacobian:
 \[\left|\frac{\partial (F_0,F_1)}{\partial (\rho_0,\Omega)}(0,\rho_0^*,\Omega_g(\rho_0^*))\right|=
 2\rho_0^*(a_{1001}+2a_{1010}-a_{0000}+\cO(\rho_0^{*2})),\ \]
 is nonzero because $\rho_0^*>0$ and
 \begin{equation}\label{a-hyp}
 a_{1001}+2a_{1010}-a_{0000}+\cO(\rho_0^{*2}))<0
 \end{equation}
since $\rho_0^*$ solves \eqref{rhocrit-eqn1} and
$\Omega_1-\Omega_0>0.$ Therefore, by the implicit function theorem,
for small $\rho_1>0,$ there is a unique solution of the system
\eqref{f4-Pzero-eqn}-\eqref{f4-Pone-eqn}:
 \ba
 \rho_0=\rho_0(\rho_1)&=&\rho_0^*+\frac{\rho_1^2}{2\rho_0^*}\left(\frac{a_{0110}+2a_{0011}-a_{1111}}{a_{1001}+2a_{1010}-a_{0000}}+\cO(\rho_0^{*2})\right)+\cO(\rho_1^4)\label{rho0-eqn}\\
 \Omega=\Omega_{asym}(\rho_1)&=&\Omega_g(\rho_0^*)+\rho_1^2\left(a_{1111}+(2a_{1010}+a_{1001})\frac{a_{0110}+2a_{0011}-a_{1111}}{a_{1001}+2a_{1010}-a_{0000}}+\cO(\rho_0^{*2})\right)+\cO(\rho_1^4),\nn\\
 &&\label{omegagg-eqn}
 \ea

\begin{rmk}
(1)  Due to equivalence of $\cN$ and $\rho_0^2+\rho_1^2$ as
parameters, for small amplitude,  we have that symmetry is broken at
 \begin{equation}
 \cN_{cr}\ \sim\ \frac{\Omega_1-\Omega_0}{|a_{0000}-a_{1001}-2a_{1010}|}
 \label{Ncr}
 \end{equation}
 (2) Note also that we have the family  of solutions
  \begin{equation}
e^{i\theta}\left(\rho_0(\rho_1)\vzero\pm\rho_1\vone+\eta(\rho_0(\rho_1),\pm\rho_1,\Omega_{asym}(\rho_1))\right),\qquad
 0\le\theta <2\pi,\ \rho_1>0.\label{sbgs}
 \end{equation}
Here the $\pm$ is present because the phase difference $\dph $
between $c_0$ and $c_1$ can be $0$ or $\pi ,$ see \eqref{dphase-eqn}
and immediately below it.
 Because $\rho_0\not=0\not=\rho_1$ this branch is neither
symmetric nor anti-symmetric. Thus, symmetry breaking has taken
place.  In the case of the double well, the $\pm$ sign in
\eqref{sbgs} shows that the bound states on this asymmetric branch
tend to localize in one of the two wells but not symmetrically in
both; see also, \cite{AFGST:02}, \cite{MKR:02}, \cite{JW:04},.....
\end{rmk}
\bigskip

\nit{\bf Case 2:}\
$\dph\in\left\{\frac{\pi}{2},\frac{3\pi}{2}\right\}:$

\nit In both cases the system
\eqref{f3-Pzero-eqn}-\eqref{f3-Pone-eqn} is equivalent to the same
system of two real equations, depending on three real parameters $\rho_0\ge
0,\ \rho_1\ge 0,\ \Omega :$
 \begin{align}
 &F_0(\rho_0,\rho_1,\Omega)\stackrel{def}{=}\Omega_0-\Omega+ a_{0000}\rho_0^2 +
 a_{0110}\rho_1^2
 +f_0(\rho_0,\rho_1 ,\Omega)\ =\ 0\label{f5-Pzero-eqn}\\
 &F_1(\rho_0,\rho_1,\Omega)\stackrel{def}{=}\Omega_1-\Omega+ a_{1111}\rho_1^2 +
 a_{1001}\rho_0^2
 +f_1(\rho_0,\rho_1 ,\Omega)\ =\ 0\label{f5-Pone-eqn}
 \end{align}
As before, in order to have another bifurcation of the symmetric
branch it is necessary to find a point,
$(\ \rho_0^{**},0,\Omega_g(\rho_0^{**})\ )$, for which:
 \begin{equation}\label{rhocrit1-eqn}
 F_1(\rho_0^{**},0,\Omega_g(\rho_0^{**})\ =\ \Omega_1-\Omega_0 +
 \left(a_{1001}-a_{0000}\right)\rho_0^{**2}+\cO(\rho_0^{**4}) \ =\
 0.
 \end{equation}
If such a point would exist we will have $\rho_0^{**}>\rho_0^{*}$
because $a_{1001}-a_{0000} > 2a_{1010}+a_{1001}-a_{0000}$ due to
$a_{1010}<0.$ Hence this bifurcation would occur later along the
symmetric branch compared to the one obtained in the previous case.
Consequently the new branch will be unstable because, as we shall
see in the next section, it bifurcates from a point where the $L_+$
operator already has two negative eigenvalues.

Moreover, it is often the case (see also the numerical results of
section \ref{sec:numerics}) that the equation \eqref{rhocrit1-eqn} has
no solution due to the wrong sign of the dominant coefficient, i.e.
$a_{1001}-a_{0000} > 0.$  This can be easily checked, in particular,
e.g., for $g=-1$ and large separation between the potential wells, using
(\ref{psilargeL}).

 \section{Exchange of stability at the bifurcation point}

In this section we consider the dynamic stability of the symmetric
and asymmetric waves, associated with the branch bifurcating from the zero state at the {\bf ground state frequency}, $\Omega_0$,  of the linear Schr\"odinger operator $-\Delta +V(x)$; see figure
 \ref{fig:gen_bifdiag}. The notion of stability
with which we work is $H^1$ - orbital Lyapunov stability.
 \begin{defin} The family of nonlinear bound states $\{\Psi_\Omega\
e^{-i\Omega t} : \theta\in[0,2\pi)\ \}$ is $H^1$~-~orbitally
Lyapunov stable if for every $\varepsilon>0$ there is a
$\delta(\varepsilon)>0$, such that if the initial data $u_0$
satisfies
 \begin{equation} \inf_{\theta\in[0,2\pi)} \|
u_0(\cdot)\ -\
\Psi_\Omega(\cdot)e^{i\theta}\|_{H^1}\ <\
\delta\ \ \ ,\nn
 \end{equation} then for all
$t\ne0$, the solution $u(x,t)$ satisfies
\begin{equation} \inf_{\theta\in[0,2\pi)} \|
u(\cdot,t)\ -\
\Psi_\Omega(\cdot)e^{i\theta}\|_{H^1}\ <\
\varepsilon.\nn
 \end{equation}
 \end{defin} In this section we  prove the following theorem:

 \begin{theo}\label{theo:exchange-stab}  The symmetric branch is
$H^1$ orbitally Lyapunov stable for $0\le\rho_0 <\rho_0^*$, or
equivalently $0<\cN<\cN_{cr}$. At the bifurcation point
$\rho_0=\rho_0^*\ (\cN=\cN_{cr})$, there is a exchange of stability
from the symmetric branch to the asymmetric branch. In particular,
for $\cN>\cN_{cr}$ the asymmetric state is stable and the symmetric
state is unstable.
 \end{theo}

We summarize basic results on stability and instability.
  Introduce $L_+$ and $L_-$,  real and imaginary parts, respectively, of the  linearized operators  about $\Psi_\Omega$:
  \ba
L_+\ =\  L_+[\Psi_\Omega]\cdot&=&(H-\Omega)\cdot\ +\ 
 \D_uN(u,u,u)\left.\right|_{\Psi_\Omega}\cdot\nn\\
 &\equiv&(H-\Omega)\cdot+D_uN[\Psi_\Omega](\cdot)\label{lp-def}\nn\\
L_-\ =\  L_-[\Psi_\Omega]\cdot&=&(H-\Omega)\cdot+
  N(\Psi_\Omega ,\Psi_\Omega,\Psi_\Omega)(\Psi_\Omega)^{-1}\ \cdot\label{lm-def}
 \ea
By (\ref{bs-eqn}) and (\ref{Nphi123def}), $L_-\Psi_\Omega=0$. 
 
     We state a special case of known results on stability and instability,
     directly applicable to the symmetric branch which bifurcates from
      the zero state at the ground state frequency of $-\Delta+V$.
  \begin{theo}\label{theo:stab-theory} \cite{We:85,We:86, GSS}
  \begin{itemize}
  \item[(1)]\ ({\it Stability})\ Suppose $L_+$ has exactly one negative eigenvalue and $L_-$ is non-negative. Assume that
  \begin{equation}
  \frac{d}{d\Omega}\ \int |\Psi_\Omega(x)|^2 dx\ <\ 0
  \label{slope}
  \end{equation}
  Then, $\Psi_\Omega$ is $H^1$ orbitally stable.
  \item[(2)]\ ({\it Instability})\  Suppose $L_-$ is non-negative.
    If $n_-(L_+)\ge2$ then the linearized dynamics about $\Psi_\Omega$ has spatially localized solution which is exponentially growing in time. Moreover, $\Psi_\Omega$ is not $H^1$ orbitally stable.
  \end{itemize}
  \end{theo}

First we claim that along the  branch of symmetric solutions,
bifurcating from the zero solution at frequency $\Omega_0$, the
hypothesis on $L_-$ holds. To see that the operator
$L_-[\Psi_\Omega]$ is always non-negative, consider
$L_-[\Psi_{\Omega_0}]=L_-[0]=-\Delta+V-\Omega_0$. Clearly, $L_-[0]$
is a non-negative operator because $\Omega_0$ is the lowest
eigenvalue of $-\Delta+V$.
 Since clearly we have $L_-\Psi_\Omega=0$,
 $0\in spec(L_-[\Psi_\Omega])$.  Since the lowest eigenvalue is necessarily simple,  by continuity there cannot be
any negative eigenvalues for $\Omega$ sufficiently close to $\Omega_0$. Finally,  if for some $\Omega$, $L_-$ has a negative eigevalue, then by continuity there would be an $\Omega_*$
for which $L_-[\Psi_{\Omega_*}$  would have a double eigenvalue at zero and no negative spectrum. But this contradicts that the ground state is simple.
Therefore, it is the quantity $n_-(L_+)$, which controls whether or not $\Psi_\Omega$ is stable.

Next we discuss the slope condition (\ref{slope}). It is clear from the
construction of the branch $\Omega\mapsto \Psi_\Omega$ that
 (\ref{slope}) holds for $\Omega$ near $\Omega_0$. Suppose now
 that $\D_\Omega \int |\Psi_\Omega|^2 =0$. Then, $\langle \Psi_\Omega, \D_\Omega\Psi_\Omega\rangle=0$.
As shown below,  $L_+$ has exactly one negative eigenvalue for
$\Omega$ sufficiently near $\Omega_0$. It follows that $L_+\ge0$ on $\{\Psi_\Omega\}^\perp$ \cite{We:85, We:86}. Therefore,  we have
$(L_+^{1\over2}\D_\Omega\Psi_\Omega,L_+^{1\over2}\D_\Omega\Psi_\Omega)=(L_+\D_\Omega\Psi_\Omega,\D_\Omega\Psi_\Omega)=(\Psi_\Omega,\D_\Omega\Psi_\Omega)=0$. Therefore, $L_+^{1\over2}\D_\Omega\Psi_\Omega=0$, implying $\Psi_\Omega=L_+\D_\Omega\Psi_\Omega=0$, which is a contradiction. It follows that (\ref{slope}) holds so long as $L_+>0$ on $\{\Psi_\Omega\}^\perp$ and when (\ref{slope}) first fails, it does so due to a non-trivial element of the nullspace of $L_+$.

Therefore $\Psi_\Omega$ is stable so long as $n_-(L_+)$ does not increase. We shall now show
that for $\cN<\cN_{cr}$, $n_-(L_+[\Psi_\Omega])=1$ but that along the symmetric branch for $\cN>\cN_{cr}$
$n_-(L_+[\Psi_\Omega])=2$. Furthermore, we show that along the bifurcating asymmetric branch, the hypotheses of Theorem \ref{theo:stab-theory} ensuring stability hold.

 \begin{rmk}\label{rmk:index-jump}
  For simplicity we have considered the most important case, where there is a transition from dynamical stability to dynamical instability along the symmetric branch, bifurcating from the ground state of $H$.
  However,  our analysis which actually shows that along any symmetric branch, associated with any of the eigenvalues, $\Omega_{2j}, j\ge0$ of $H$, there is a critical $\cN=\cN_{cr}(j)$, such that as $\cN$ is increased through $\cN_{cr}(j)$,  $n_-(L_+^{(j)})$ the number of negative eigenvalues of the linearization about the symmetric state along the $j^{th}$ symmetric branch increases by one.  By the results in \cite{Jones,ManosG,kapitula:04},  this has implications for the number of unstable modes of higher order ($j\ge1$) symmetric states.
  \end{rmk}

 Consider the spectral problem for $L_+=L_+[\Psi_\Omega]$:
 \be\label{lp-eval}
 L_+[\Psi_\Omega]\phi=\mu\phi
 \ee
We now formulate a Lyapunov-Schmidt reduction of (\ref{lp-eval}) and
then relate it to our formulation for nonlinear bound states.
We first decompose $\phi$ relative to the states $\vzero,\ \vone$ and their
orthogonal complement:
\begin{equation}
\phi\ =\ \alpha_0\psi_0\ +\ \alpha_1\psi_1\ +\ \xi,\ \ \
\left(\psi_j,\xi\right)=0,\ j=0,1 \nn\end{equation} Projecting
\eqref{lp-eval} onto $\vzero,\ \vone$ and onto the range of $\Pc$ we
obtain the system:
 \ba
 \la\ \vzero ,L_+[\Psi_\Omega](\alpha_0\psi_0\ +\ \alpha_1\psi_1\ +\ \xi)\ \ra&=&\mu\alpha_0 \label{lp-zero}\\
 \la\ \vone ,L_+[\Psi_\Omega](\alpha_0\psi_0\ +\ \alpha_1\psi_1\ +\ \xi)\ \ra&=&\mu\alpha_1 \label{lp-one}\\
 (H-\Omega)\xi+D_uN[\Psi_\Omega](\alpha_0\vzero+\alpha_1\vone+\xi)&=&\mu\xi .\label{lp-cont}
 \ea
The last equation can be rewritten in the form:
 \be\label{xi-eqn}
\left[\ I\ +\ (H-\Omega-\mu)^{-1}\Pc D_uN[\Psi_\Omega]\ \right]\xi\
 =\ -(H-\Omega-\mu)^{-1}\Pc D_uN[\Psi_\Omega](\alpha_0\vzero+\alpha_1\vone)
 \ee
 The operator on the right hand side of (\ref{xi-eqn}) is essentially the Jacobian studied in the proof of
 Proposition
\ref{prop:eta}, evaluated at $\Omega+\mu$.
 Hence, by the proof of
  Proposition \ref{prop:eta}, if $\Omega+\mu$ satisfies (\ref{Omegad})
   and
$\|\Psi_\Omega\|_{H^2}\le \cN_*$, then the operator
$I+(H-\Omega-\mu)^{-1}\Pc D_uN[\Psi_\Omega]$ is invertible on $H^2$
and \eqref{xi-eqn} has a unique solution
 \ba
 \xi&\stackrel{def}{=}&\ \xi[\mu,\alpha_0,\alpha_1,\Omega]\nn\\
& \equiv&\
  Q[\mu,\Psi_\Omega](\alpha_0\vzero+\alpha_1\vone) \label{xi-sln}\\
 &=&-(I+(H-\Omega-\mu)^{-1}\Pc D_uN[\Psi_\Omega])^{-1}(H-\Omega-\mu)^{-1}\Pc D_uN[\Psi_\Omega](\alpha_0\vzero+\alpha_1\vone)\nn\\
 &=& \cO\left[(|\rho_0|+|\rho_1|)^2\right]\ \left[\ \alpha_0\psi_0+\alpha_1\psi_1\right].\nn
 \ea
The last relation follows from $D_uN[\psi]$ being a quadratic form
in
$\Psi_\Omega=\rho_0\vzero+\rho_1\vone+\cO((|\rho_0|+|\rho_1|)^3).$

Substitution of the expression for $\xi$ as a functional of
$\alpha_j$ into \eqref{lp-zero} and \eqref{lp-one} we get a closed
system of two real equations:
 \ba\label{lp-red}
 (\Omega_0-\Omega)\alpha_0\ +\ \la\psi_0,D_uN[\Psi_\Omega]\  (I+Q[\mu ,\Psi_\Omega])\ (\alpha_0\vzero+\alpha_1\vone)\ra\ =\ \mu\ \alpha_0 \nn\\
  (\Omega_1-\Omega)\alpha_1\ +\ \la\psi_1,D_uN[\Psi_\Omega]\  (I+Q[\mu ,\Psi_\Omega ])\ (\alpha_0\vzero+\alpha_1\vone)\ra\ =\ \mu\ \alpha_1
 \ea
 The system (\ref{lp-red}) is the Lyapunov Schmidt reduction of the linear eigenvalue problem for $L_+$ with eigenvalue parameter $\mu$. {\it Our next step will be to write it in a form, relating it to the linearization of the Lyapunov Schmidt reduction of the nonlinear problem.}
 \begin{rmk} For $\|\Psi_\Omega\|_{H^2}\le n_*,$ the above system is equivalent to the eigenvalue
problem for the operator $L_+[\Psi_\Omega]$ with eigenvalue
parameter $\mu$ as long as \ref{Omegad}) holds with $\Omega$ replaced by $\Omega+\mu$.
This restriction on the spectral parameter, $\mu$, is in fact very mild and has no impact on the analysis.  This is because we are primarily interested in $\mu$ near zero, as we are are interested in detecting the crossing of an eigenvalue of $L_+$ from positive to negative reals
 as $\cN$ is increased beyond some $\cN_{cr}$.    Values of $\mu$ for which (\ref{Omegad}) does not hold,  do not play a role in any change of index, $n_-(L_+).$
\end{rmk}

First rewrite (\ref{lp-red}) as
\ba
& (\Omega_0-\Omega-\mu)\alpha_0\ +\ \la\psi_0,D_uN[\Psi_\Omega]\
  (I+Q[0 ,\Psi_\Omega])\ (\alpha_0\vzero+\alpha_1\vone)\ra
  \label{lp-red1a}\\
  &\ \ +\
  \la\psi_0,D_uN[\Psi_\Omega]\
  \Delta Q[\mu ,\Psi_\Omega]\ (\alpha_0\vzero+\alpha_1\vone)\ra
    =\ 0\nn\\
&  (\Omega_1-\Omega-\mu)\alpha_1\ +\ \la\psi_1,D_uN[\Psi_\Omega]\  (I+Q[0 ,\Psi_\Omega ])\ (\alpha_0\vzero+\alpha_1\vone)\ra\nn\\
&\ \ +\
  \la\psi_1,D_uN[\Psi_\Omega]\
  \Delta Q[\mu ,\Psi_\Omega]\ (\alpha_0\vzero+\alpha_1\vone)\ra
    =\ 0. \label{lp-red1b}
 \ea
 Here,
 \begin{equation}
 \Delta Q\left[\mu,\Psi_\Omega\right]\
  =\ Q\left[\mu,\Psi_\Omega\right]\ -\
  Q\left[0,\Psi_\Omega\right].
  \label{DeltaQ}
  \end{equation}
  Note that
  terms involving $\Delta Q$ in
  (\ref{lp-red1a},\ref{lp-red1b}) are of size $\cO[ (\rho_0^2+\rho_1^2)\mu\alpha_j]$.

 \begin{prop}\label{prop:Qprop1}
  \be
Q[0,\Psi_\Omega](\alpha_0\psi_0+\alpha_1\psi_1)\ =\
\D_{\rho_0}\eta[\rho_0,\rho_1,\Omega]\ \alpha_0\ +\
\D_{\rho_1}\eta[\rho_0,\rho_1,\Omega]\ \alpha_1 \label{Qeta}\ee
\end{prop}
\nit{\it Proof of Proposition \ref{prop:Qprop1}:}\
Recall that $\eta$ satisfies
\begin{align}
&F(\rho_0,\rho_1,\Omega,\eta)\ \equiv\ \eta+(H-\Omega)^{-1}\Pc
 N[\ \rho_0\vzero+\rho_1\vone+\eta\ ]\ =\ 0,
 \label{Feq0}
 \end{align}

\nit  Differentiation with respect to $\rho_j,\ j=0,1$ yields
 \begin{equation}
 \left(\ I\ +\ (H-\Omega)^{-1}\Pc D_uN[\Psi_\Omega]\ \right)\ \D_{\rho_j}\eta\ = \ -\left(\ H-\Omega\ \right)^{-1}\Pc D_u N[\Psi_\Omega]\psi_j,
 \label{Qprop1-j}
 \end{equation}
where
\begin{equation}
  \Psi_\Omega=\rho_0\vzero+\rho_1\vone+\eta[\rho_0,\rho_1,\Omega].
  \nn\end{equation}
 Thus,
 \begin{equation}
 \D_{\rho_j}\ \eta\ =\ Q[0,\Psi_\Omega]\ \psi_j,
 \label{Drhojj-eta}
 \end{equation}
from which Proposition \ref{prop:Qprop1} follows.

We now use Proposition \ref{prop:Qprop1} to rewrite the first inner products in equations (\ref{lp-red1a})-(\ref{lp-red1b}). For $k=0,1$ 
\begin{align}
&\la\psi_k,D_uN[\Psi_\Omega]\
  (I+Q[0 ,\Psi_\Omega])\ (\alpha_0\vzero+\alpha_1\vone)\ra\nn\\
  &= \sum_{j=0}^1 \la\psi_k,D_uN[\rho_0\psi_0+\rho_1\psi_1+\eta](\psi_j+\D_{\rho_j}\eta)\ra\alpha_j\nn\\
  &=\ \sum_{j=0}^1\ \frac{\D}{\D\rho_j}\la\psi_k,N[\Psi_\Omega]\ra\ \alpha_j\nn\\
  &=\ \sum_{j=0}^1\ \D_{\rho_j}\ \la\psi_k,N[\rho_0\psi_0+\rho_1\psi_1]\ra\ \alpha_j\ +\ \D_{\rho_j}\ \left[\ \rho_k f_k(\rho_0,\rho_1,\Omega)\ \right],
  \label{rewrite}
  \end{align}
where $N[\psi_\Omega]=N[\rho_0\psi_0+\rho_1\psi_1]+\cR$; see
 equations (\ref{Pzero-eqn}-\ref{Pc-eqn}), (\ref{f0}-\ref{f1}).
Therefore, the Lyapunov-Schmidt reduction of the eigenvalue problem for $L_+$ becomes
\ba
& (\Omega_0-\Omega-\mu)\alpha_0\ +\
\sum_{j=0}^1\ \D_{\rho_j}\ \la\psi_0,N[\rho_0\psi_0+\rho_1\psi_1] \ra\ \alpha_j\ +\ \D_{\rho_j}\ \left[\ \rho_0 f_0(\rho_0,\rho_1,\Omega)\ \right]  \label{lp-red2a}\\
  &\ \ +\
  \la\psi_0,D_uN[\Psi_\Omega]\
  \Delta Q[\mu ,\Psi_\Omega]\ (\alpha_0\vzero+\alpha_1\vone)\ra
    =\ 0\nn\\
&  (\Omega_1-\Omega-\mu)\alpha_1\ +\
\sum_{j=0}^1\ \D_{\rho_j}\ \la\psi_1,N[\rho_0\psi_0+\rho_1\psi_1]\ \ra\ \alpha_j\ +\ \D_{\rho_j}\ \left[\ \rho_1 f_1(\rho_0,\rho_1,\Omega)\ \right]
\nn\\
&\ \ +\
  \la\psi_1,D_uN[\Psi_\Omega]\
  \Delta Q[\mu ,\Psi_\Omega]\ (\alpha_0\vzero+\alpha_1\vone)\ra
    =\ 0. \label{lp-red2b}
 \ea

 This can be written succinctly in matrix form as
 \be\label{e-sys}
\left[\  M \ -\mu\ +\ {\cal C}(\mu)\ \right]\
\left(\begin{array}{l}\alpha_0 \\ \alpha_1\end{array}\right) =\
\left(\begin{array}{l} 0 \\ 0\end{array}\right),
 \ee
where
{\footnotesize{
\begin{align}
 &M=M[\Omega,\rho_0,\rho_1]\nn\\
 &\left(\begin{array}{rl}
         \Omega_0-\Omega+3a_{0000}\rho_0^2+(a_{0110}+2a_{0011})\rho_1^2+\D_{\rho_0}(\rho_0f_0)&
         2(a_{0110}+2a_{0011})\rho_0\rho_1+\D_{\rho_1}(\rho_0f_0)\nn\\
         2(2a_{1010}+a_{1001})\rho_0\rho_1+\D_{\rho_0}(\rho_1f_1)         & (\Omega_1-\Omega)+ 3a_{1111}\rho_1^2 +
         (2a_{1010}+a_{1001})\rho_0^2
         +\D_{\rho_1}(\rho_1f_1)
         \end{array}\right)\nn\\
         &\label{M-def}
 \end{align}
 }}
and
 \be\label{C-def}
 {\cal C}(\mu)_{lm}=\la\psi_l,D_uN[\Psi_\Omega]\Delta
 Q[\mu,\Psi_\Omega]\psi_m\ra,\quad l,m=0,1.
 \ee
Note that
 \be\label{C-zero}
 {\cal C}(\mu=0)=0.
 \ee

Recall that $\mu$ is the spectral parameter for the eigenvalue problem $L_+$, (\ref{lp-eval}) and we are interested in $n_-(L_+[\Psi_\Omega])$, the number of negative eigenvalues along a family of nonlinear bound states $\Omega\mapsto\Psi_\Omega$. By Theorem \ref{theo:stab-theory} $n_-(L_+)$ determines the stability or instability of a particular state. This question has now been mapped to the problem of following the roots of
\be\label{D-def1}
 D(\mu,\rho_0,\rho_1)=\det(\mu I-M-{\cal C}(\mu))=0,
 \ee
where $\rho_0$ and $\rho_1$ are parameters along the different branches of nonlinear bounds states.  Since ${\cal C}(\mu)$, defined in (\ref{C-def}) is small for small amplitude nonlinear bound states, we expect the roots, $\mu$, to be small perturbations of the eigenvalues of the matrix $M$. We study these roots  along the
symmetric ($M=M(\Omega_g(\rho_0),\rho_0,0)$) and asymmetric branch  ($M=M(\Omega_{asym}(\rho_1),\rho_0(\rho_1),\rho_1)$) using the implicit function theorem.
\bigskip

 \nit\centerline{\bf Symmetric branch:}

 \nit Along the symmetric branch:
 \[
 \rho_1=0,\quad \rho_0\ge 0,\quad
 \Omega=\Omega_g=\Omega_0+a_{0000}\rho_0^2+\cO(\rho_0^4),\quad
 \Psi_\Omega=\rho_0\psi_0+\eta(\rho_0,0,\Omega)={\rm symmetric}.
 \]
Thus, $D=D(\mu,\rho_0)$. Moreover, the system \eqref{e-sys} is diagonal. This is because $Q$, and  hence
$\Delta Q$, preserve parity at a symmetric $\Psi_\Omega$; see their
definitions. 
Therefore ${\cal
C}_{01}=0={\cal C}_{10}$, each the scalar product of an even and  an
odd function. Moreover from \eqref{f0f1-ser1} we get:
$\frac{\partial f_j}{\partial\rho_1}(\rho_0,0,\Omega)=0,\ j=0,1.$

Therefore, the matrix $\mu I-M-{\cal C}(\mu)$ is diagonal and $\mu$ is an eigenvalue of $L_+[\psi_{\Omega_g(\rho_0)}]$ if and only if $\mu$ is a root of either
 \be
 P_0(\mu,\rho_0)\ \equiv\ \mu-M_{00}(\rho_0)-{\cal C}_{00}(\mu,\rho_0)=0\label{mu0-eqn}\ee
 or
 \be
 P_1(\mu,\rho_0)\ \equiv\ \mu-M_{11}(\rho_0)-{\cal
 C}_{11}(\mu,\rho_0)=0\label{mu1-eqn}
 \ee
Both $P_0$ and $P_1$ are analytic in $\mu$ and $\rho_0$ and it is
easy to check that
 \be
 P_0(0,0)\ =0,\ \ \ \
 \partial_\mu P_0(0,0)=1\nn
 \ee
and
 \be
 P_1(\Omega_1-\Omega_0,0)=0,\ \ \ \
 \partial_\mu P_1(\Omega_1-\Omega_0,0)=1.\nn
 \ee
 Formally differentiating (\ref{mu0-eqn}) or (\ref{mu1-eqn}) with respect to $\rho_0$ gives:
 \be
\D_{\rho_0} \mu_j=\frac{\D_{\rho_0}M_{jj}+\D_{\rho_0}C_{jj}}{1-\partial_\mu C_{jj}}.
 \label{Dmu}
 \ee
 By the implicit function theorem (\ref{mu0-eqn}) and  (\ref{mu1-eqn}) define, respectively, $\mu_0$ and $\mu_1$ as smooth functions of $\rho$ provided
 \be\label{Cinv-cond}|\partial_\mu
 {\cal C}_{jj}|<1,\qquad j=0,1\ee
 A direct calculation using \eqref{xi-sln} and estimates
\eqref{N-est}, \eqref{res-ubd} shows that in the regime of interest:
$\Omega$ satisfying (\ref{Omegad}),  it suffices to have
 \be\label{nbs-cond}
 \|\Psi_\Omega\|_{H^2}\le
 n_*\left(9\max (\|\psi_0\|_{H^2},\|\psi_1\|_{H^2})\right)^{-\frac{1}{4}}
 \ee
where $n_*$ is given by Proposition \ref{prop:eta}. The latter can be
reduced to an estimate on $\rho_0$ via the above definition of
$\Psi_\Omega $ and \eqref{eta-est1} as in the end of the proof of
Proposition \ref{prop:eta}.

Therefore, under conditions \eqref{Omegad} and \eqref{nbs-cond}, we have
a unique solution $\mu_0,$ respectively $\mu_1,$ of \eqref{mu0-eqn},
respectively \eqref{mu1-eqn}. Moreover, the two solutions are
analytic in $\rho_0$ and, for small $\rho_0,$ we have the following
estimates:
 \ba
 \mu_0&=&2a_{0000}\rho_0^2+\cO(\rho_0^4)<0\label{e0-eqn}\\
\mu_1&=&\Omega_1-\Omega_0+\cO(\rho_0^2)>0,\label{e1-eqn}
 \ea
where we used $a_{0000}\equiv  g\langle \psi_0^2,K[\psi_0^2]\rangle<0,$ and
$\mu_1(\rho_0=0)=\Omega_1-\Omega_0>0.$

We claim that $\mu_1$ changes sign for the first time at
$\rho_0=\rho_0^*.$ Indeed, by continuity, the sign can only change
when $\mu_1=0,$ {\it  i.e.} when \eqref{mu1-eqn} has a solution of the form
$(0,\rho_0).$ Since ${\cal C}_{11}(0,\rho_0)=0,$ see \eqref{C-zero},
\eqref{mu1-eqn} becomes
 \[0=M_{11}(\rho_0)=\Omega_1-\Omega_g(\rho_0)+(2a_{1010}+a_{1001})\rho_0^2+f_1(\rho_0,0,\Omega_g)\ =\ F_1(\rho_0,0,\Omega_g(\rho_0));\]
see \eqref{M-def} and note that $\rho_1=0.$  But this equation is
the same as \eqref{rhocrit-eqn}, which determines $\rho_0^*$, then bifurcation point. Thus, as expected, $D(\mu,\rho_0)=0$ has a root
 $\rho_1(\rho_0^*)=0$ or equivalently $L_+$ has a zero eigenvalue at the bifurcation point. Note that the associated null eigenfunction has odd parity  in one space dimension, and is more generally, non-symmetric and changes sign.

To see that $\mu_1(\rho_0)$ changes sign at $\rho_0=\rho_0^*$
we differentiate \eqref{mu1-eqn} with respect to $\rho_0$ at
$\rho_0=\rho_0^*$ and obtain from (\ref{Dmu}) that
 \[\D_{\rho_0}\mu_1=\frac{\D_{\rho_0}M_{11}+\D_{\rho_0}C_{11}}{1-\partial_\mu C_{11}}<0.\]
This follows because  the denominator is positive, by
\eqref{Cinv-cond},  while direct calculation gives for the
numerator: \be \D_{\rho_0}
M_{11}(\rho_0^*)+\D_{\rho_0}C_{11}(\rho_0^*)=2\rho_0^*\left(a_{1001}+2a_{1010}-a_{0000}+\cO(\rho_0^{*2})\right)<0\nn
\ee see \eqref{a-hyp}. Therefore $\mu_1$ becomes negative for
$\rho_0>\rho_0^*$ at least when $|\rho_0-\rho_0^*|$ is small enough.

In conclusion, $L_+[\Omega_g(\rho_0)]$ has exactly one negative eigenvalue for
$0\le\rho_0<\rho_0^*$ and two negative eigenvalues for
$\rho_0>\rho_0^*$ and $|\rho_0-\rho_0^*|$ small. Therefore, following
the criteria of  \cite{We:85,We:86, GSS, Jones,ManosG,Kapitula}, the
symmetric branch is stable for $0\le\rho_0<\rho_0^*$ and becomes
unstable past the bifurcation point.
\bigskip

\nit\centerline {\bf Asymmetric branch: Stability for
$\cN>\cN_{cr}$}

\nit Finally, we study the behavior of eigenvalue problem \eqref{e-sys}
on the asymmetric branch:
 \ba
 0\le\rho_1\ll  1 & & \nn\\
 \rho_0=\rho_0(\rho_1)&=&\rho_0^*+\frac{\rho_1^2}{2\rho_0^*}\left(\frac{a_{0110}+2a_{0011}-a_{1111}}{a_{1001}+2a_{1010}-a_{0000}}+\cO(\rho_0^{*2})\right)+\cO(\rho_1^4)\label{r0as}\\
 \Omega=\Omega_{asym}(\rho_1)&=&\Omega_g(\rho_0^*)+\rho_1^2\left(a_{1111}+(2a_{1010}+a_{1001})\frac{a_{0110}+2a_{0011}-a_{1111}}{a_{1001}+2a_{1010}-a_{0000}}+\cO(\rho_0^{*2})\right)\nn\\
 &&\ +\cO(\rho_1^4),\label{Omega-asym}\\
 \Psi_\Omega&=&\rho_0(\rho_1)\vzero+\rho_1\vone+\eta(\rho_0(\rho_1),\rho_1,\Omega_{asym}(\rho_1))\nn
 \ea
The eigenvalues will be given by the zeros of the real valued
function
 \be\label{D-def}
 D(\mu,\rho_1)=\det(\mu I-M(\rho_1)-{\cal C}(\mu,\rho_1)),
 \ee
which is analytic in $\mu$ and $\rho_1$ for $\Omega+\mu$ satisfying  (\ref{Omegad}) and
$\rho_1$ small. Note that at $\rho_1=0$ we are still on the
symmetric branch at the bifurcation point $\rho_0=\rho_0^*.$ Hence,
the matrix is diagonal and
 \be\label{D0-eqn}
 D(\mu,0)=P_0(\mu,\rho_0^*)P_1(\mu,\rho_0^*),
 \ee
where $P_j,\ j=0,1$ are defined in \eqref{mu0-eqn}-\eqref{mu1-eqn}.
In the previous subsection we showed that each $P_j(\cdot,\rho_0^*)$
has exactly one zero, $\mu_j ,$ on the interval
$-\infty<\mu<\-d_*-\Omega_g(\rho_0^*)>0.$ The zeros were simple, by our implicit function theorem application in which,
 \be\label{der-est}\partial_\mu P_j(\mu_j,\rho_0^*)=1-\partial_\mu
 C_{jj}>0,\ee
see \eqref{Cinv-cond}. In addition one can easily deduce that
$\lim_{\mu\rightarrow -\infty}P_j(\mu,\rho_0^*)=-\infty$ by using
the definitions \eqref{C-def}, \eqref{DeltaQ} and the fact that
$\|(H-\Omega-\mu)^{-1}\|_{L^2\rightarrow
H^2}\stackrel{\mu\rightarrow -\infty}{\rightarrow}0$ which implies
$\|Q[\mu,\Psi_\Omega]\|_{H^2\rightarrow H^2}\stackrel{\mu\rightarrow
-\infty}{\rightarrow}0.$

Consequently $D(\cdot,0)$ has exactly two simple zeros $\mu_0<0$ and
$\mu_1=0$ on the interval $-\infty<\mu\le
(-d_*-\Omega_g(\rho_0^*))/2>0,$ which are both simple and
$\lim_{\mu\rightarrow -\infty}D(\mu,0)=\infty$. It is well known,
and a consequence of continuity arguments and of the implicit function
theorem, that the previous statement is stable with respect to small
perturbations. More precisely, there exists $\veps>0$ such that
whenever $|\rho_1|<\veps,$ $D(\cdot,\rho_1)$ has exactly two zeros
$\mu_0(\rho_1)<0$ and $\mu_1(\rho_1)$ on the interval
$-\infty<\mu\le (-d_*-\Omega_g(\rho_0^*))/2>0,$ which are both
simple and analytic in $\rho_1.$

Since we are interested in $n_-(L_+)$, the number of negative eigenvalues of $L_+$,  we still need to determine the
sign of $\mu_1(\rho_1).$ In what follows we will show that its
derivatives satisfy
 \be\label{mu1-taylor}
 \D_{\rho_1}\mu_1(0)=0,\quad \D_{\rho_1}^2\mu_1(0)>0.
 \ee
We can then conclude that  for $0<\rho_1\ll 1,$ $\mu_1(\rho_1)>0,$ and $L_+$ has
exactly one (simple) negative eigenvalue, $\mu_0(\rho_1)$. Therefore,
the asymmetric branch is stable.

We now prove \eqref{mu1-taylor}. By differentiating
 \be\label{mu11-eqn}D(\mu_1(\rho_1),\rho_1)=0\ee
once with respect to $\rho_1$ at $\rho_1=0$ we get
 \[
\D_\mu D(0,0) \D_{\rho_1}\mu_1(0)+\partial_{\rho_1}D(0,0)=0
 \]
Using \eqref{D0-eqn} we obtain
 \be\label{D00der-est}\partial_\mu D(0,0)=P_0(0,\rho_0^*)\partial_\mu
 P_1(\mu_1=0,\rho_0^*)>0\ee
where we used \eqref{der-est}  and that $P_0(0,\rho_0^*)=-M_{00}(\rho_0^*)>0$. Using \eqref{D-def} and \eqref{C-zero} we
obtain
 \be
 \partial_{\rho_1}D(0,0)=\frac{\partial \det(M)}{\partial\rho_1}(\rho_1=0)=\det 10+\det 01,\ \ {\rm where}\label{d00}
 \ee
\bigskip

\nit $\det ij\ =$\   the determinant evaluated at $\rho_1=0$ of the matrix
obtained from $M$ by differentiating the first row $i$ times,
respectively the second row $j$ times. $\det ij$ can be evaluated using (\ref{f0f1-ser}), (\ref{rhocrit-eqn}), and (\ref{Omega-asym}).
\bigskip

\nit  Note that the second row of  $\det 10$ is zero and therefore $\det 10=0$. Furthermore, $\det 01$ is zero because its second
column is zero.
 Therefore, by (\ref{d00}) we have $\D_{\rho_1}\mu_1(0)=0.$.

 We now calculate $\D^2_{\rho_1}\mu_1(\rho_1=0)$. Differentiate
\eqref{mu11-eqn} twice with respect to $\rho_1$ at $\rho_1=0$ and
use $\D_{\rho_1}\mu_1(0)=0$ to obtain:
 \[
 \partial_\mu D(0,0)\D_{\rho_1}^2  \mu_1(0)
 +\partial_{\rho_1}^2D(0,0)=0.
 \]
which implies, by (\ref{D00der-est})
 \[ {\rm sign} (\D_{\rho_1}^2\mu_1(0))=-{\rm sign}( \partial_{\rho_1}^2D(0,0)).\]
  But, as before, \eqref{D-def} and
\eqref{C-zero} imply
 \[
 \partial_{\rho_1}^2D(0,0)=\frac{\partial^2 \det(M)}{\partial\rho_1^2}(0)=\det 20+2\det 11+\det 02<0.\]
The last inequality is a consequence of the following argument.
First, $\det 20=0$,  since its second row zero. A direct
calculation using the definition of $M$ and relations \eqref{r0as}
show:
 \ba
 \det 11&=&-4(a_{0110}+2a_{0011})(2a_{1010}+a_{1001})\rho_0^{*2}+\cO(\rho_0^{*4})\nn\\
 \det 02&=&8a_{0000}a_{1111}\rho_0^{*2}+\cO(\rho_0^{*4})\nn
 \ea
Note that in the limit of large well-separation limit ($L>>1$), all coefficients  $a_{klmn}=a_{klmn}(L)$
converge
 to the same value $g\alpha^2<0$. This implies
 \be
 2\det 11+\det 02=(-64g^2\alpha^4+\cO(e^{-\tau L}))\rho_0^{*2}+\cO(\rho_0^{*4})<0.
 \nn\ee
 Therefore, $\D_{\rho_1}^2\mu_1(0)>0$ and the proof of
\ Theorem \ref{theo:exchange-stab} is now complete.

\section{Numerical study of symmetry breaking}
\label{sec:numerics}

\nit{\bf Symmetry breaking bifurcation for fixed well-separation, $L$}

In this section we numerically compute the bifurcation diagram for the lowest energy nonlinear bound state branch for NLS-GP (\ref{nls-gp}) and compare
 these results to the predictions of the  finite dimensional
approximation Eqs. (\ref{t3-eqns}.  Specifically, we numerically compute the bifurcation structure of Eq. (\ref{nls-gp}) for a
double-well potential, $V_L(x)$,  of the form:
\begin{eqnarray}
V(x)=V_0 \left[ \frac{1}{\sqrt{4 \pi s^2}} \exp \left(- \frac{(x-L/2)^2}{4 s^2}\right) + \frac{1}{\sqrt{4 \pi s^2}} \exp
\left(- \frac{(x+L/2)^2}{4 s^2}\right) \right]. \label{poten}
\end{eqnarray}
The potential for $V_0=-1$ , $s=1$ and $L=6$ has two discrete  eigenvalues $\Omega_0=-0.1616$ and $\Omega_1=-0.12$ and
a continuous spectral part for $\Omega>0$. The linear eigenstates can also be obtained and used to
numerically compute the coefficients of the finite dimensional decomposition of Eqs. (\ref{t3-eqns}) as
$a_{0000}=-0.09397$, $a_{1111}=-0.10375$,
$a_{0011}=a_{1010}=a_{1001}=a_{0110}=-0.08836$
(for $g=-1$). Then, using (\ref{Ncrit}), we can
compute the approximate threshold in $\cN$ for bifurcation of an asymmetric branch (and the destabilization of the symmetric one):
 \begin{equation}
 \cN_{cr}\sim \cN_{cr}^{(0)}=0.24331,\ \
  \Omega_{cr}\sim \Omega_{cr}^{(0)}\equiv\Omega_0\ +\ a_{0000}\ \cN_{cr}^{(0)} =-0.18447.
  \nn\end{equation}
  We expect good agreement because the values of $s$ and $L$  suggest the regime of large $L$, where our rigorous theory holds.

Using numerical fixed-point iterations (in particular Newton's method),
we have obtain the branches of the nonlinear eigenvalue
problem (\ref{bs-eqn}). To study the stability of a solution, $u_0$,  of
(\ref{bs-eqn}), consider the evolution of a small perturbation of it:
\begin{eqnarray}
u=e^{-i \Omega t} \left[u_0(x) + \left( p(x) e^{\lambda t} +
q(x) e^{\bar{\lambda} t} \right) \right].
\label{linearization}
\end{eqnarray}
Keeping only linear terms in $p,q$, we obtain a linear evolution equation, whose normal modes satisfy a linear eigenvalue problem  with spectral parameter, which we denote by
$\lambda$ and eigenvector $(p(x),\bar{q}(x))^T$.

\begin{figure}[ht]
\begin{center}
\begin{tabular}{cc}
(a) & (b) \\
\includegraphics[height=6cm]{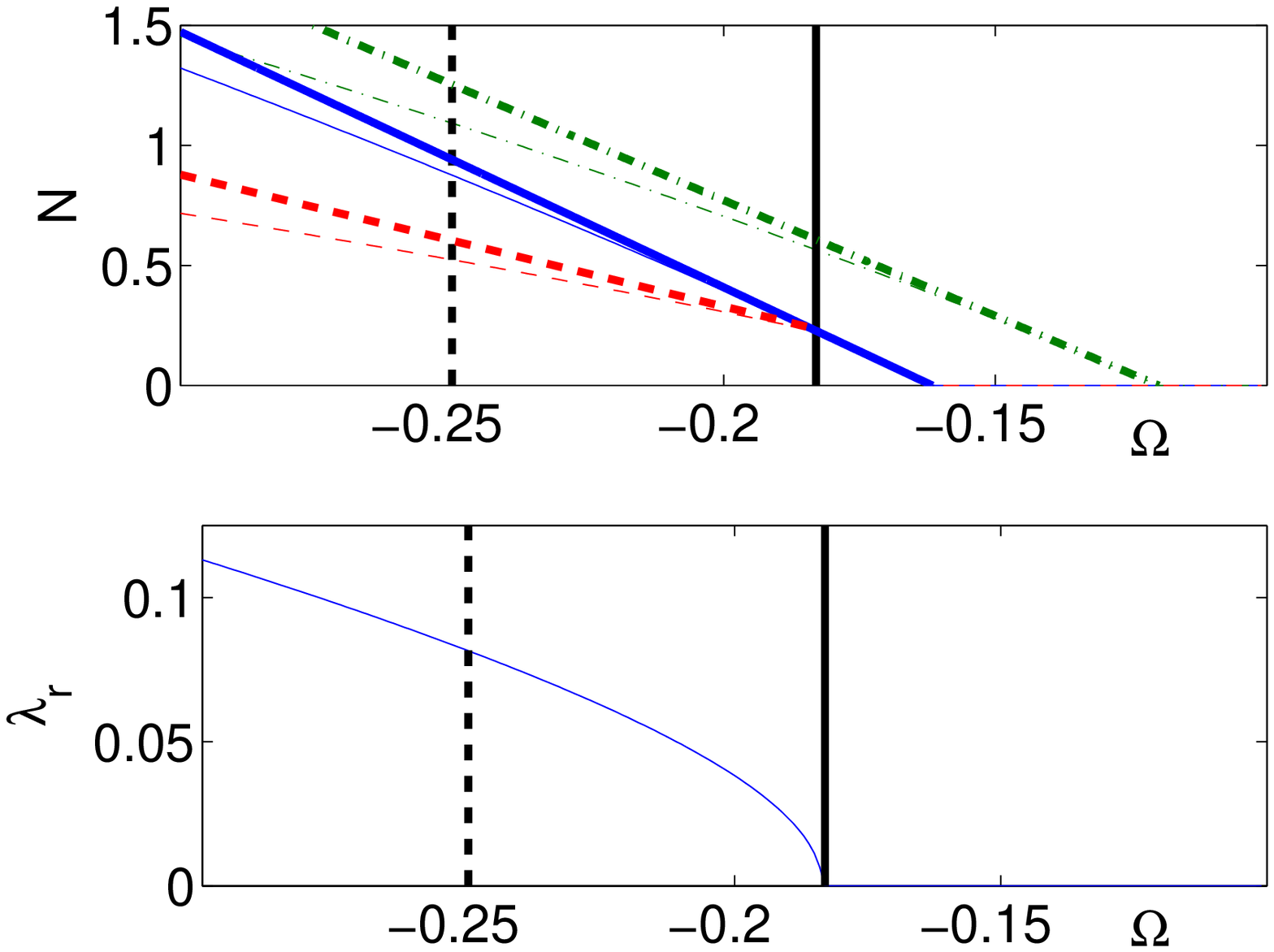} &
\includegraphics[height=6cm]{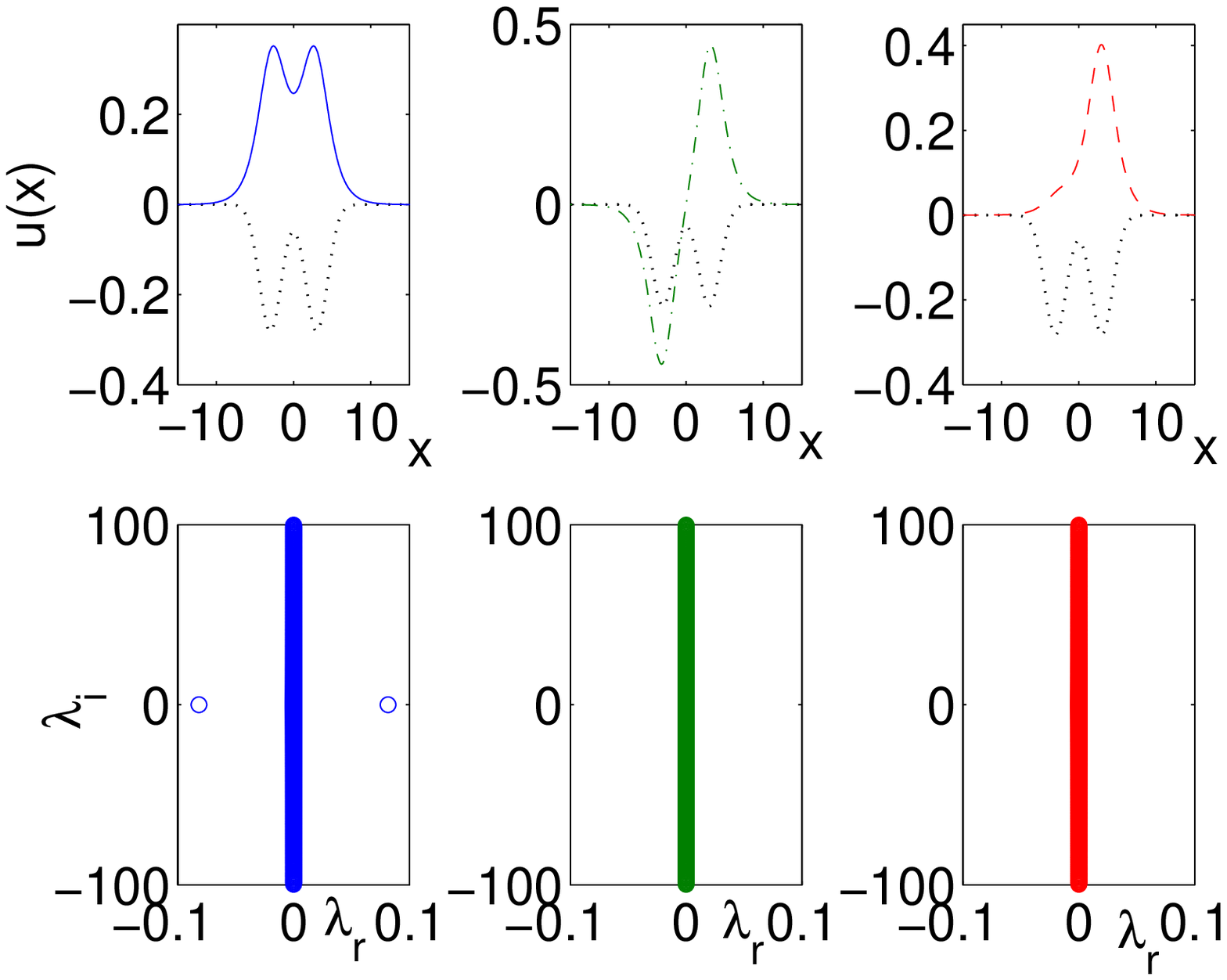} \\
\end{tabular}
\end{center}
\caption{(Color Online) The figure shows the typical numerical
bifurcation results for the cubic case and their comparison with
the finite dimensional analysis of Section \ref{sec:finite-dim-approx}.
Panel (a) shows the bifurcation diagram in the top subplot and the
relevant real eigenvalues in the bottom subplot. In the top,
the solid (blue) line represents the symmetric branch, the
dash-dotted (green) line the anti-symmetric branch, while the
dashed (red) line represents the bifurcating asymmetric branch.
The thin lines indicate the numerical findings, while the thick
ones show the corresponding
finite-dimensional, weakly nonlinear predictions. The solid vertical
(black) line indicates the critical point (of $\Omega \approx -0.1835$)
obtained numerically. The dashed vertical (black) line is a guide to
the eye for the case with $\Omega=-0.25$, whose detailed results are
shown in panel (b). The bottom subplot of panel (a) shows the real
eigenvalue (as a function of $\Omega$)
of the symmetric branch that becomes unstable for  $\Omega<-0.1835$.
Panel (b) shows using the same symbolism as panel (a) the symmetric
(left), anti-symmetric (middle) and asymmetric (right) branches
and their linearization eigenvalues (bottom subplots) for
$\Omega=-0.25$.
The potential is shown by a dotted black line.}
\label{bfig1}
\end{figure}

Our computations for the simplest case of the cubic
nonlinearity with $K[\psi \bar{\psi}]=\psi \bar{\psi}$
are shown in Figure \ref{bfig1} (for $g(x)=-1$).
In particular, the top subplot of panel (a) shows
the full numerical results by thin lines (solid for
the symmetric solution, dashed for the bifurcating asymmetric and
dash-dotted for the anti-symmetric one) and compares them
with the predictions based on the finite dimensional truncation, (\ref{t3-eqns})
shown by the corresponding thick lines. The approximate threshold values $\cN_{cr}$ and $\Omega_{cr}$ are found numerically to be
 $ \Omega_{cr}^{(0)} \approx -0.1835$, $\cN_{cr}^{(0)}
\approx 0.229$.  This suggests a relative error
in its evaluation by the finite-dimensional reduction
of less than $1 \%$. This critical point is indicated
by a solid vertical (black) line in panel (a). For
$\Omega > \Omega_{cr}^{(0)}$, there exist two branches in
the problem, namely the one that bifurcates from the
symmetric linear state (this branch exists for
$\Omega<\Omega_0$) and the one that bifurcates from
the anti-symmetric linear state (and, hence, exists
for $\Omega<\Omega_1$). For $\Omega < \Omega_{cr}^{(0)} $,
the symmetric branch becomes unstable due to a real
eigenvalue (see bottom subplot of panel (a)), signalling
the emergence of a new branch, namely the asymmetric
one. All three branches are shown for $\Omega=-0.25$
(indicated by dashed vertical (black) line in panel (a))
in panel (b) and their corresponding linearization
spectrum $(\lambda_r,\lambda_i)$ is shown for the eigenvalues
$\lambda=\lambda_r + i \lambda_i$.

\bigskip

\nit{\bf Symmetry breaking threshold, $\cN_{cr}(L)$ as $L$ varies}

\begin{figure}[ht]
\begin{center}
\begin{tabular}{cc}
(a) & (b) \\
\includegraphics[height=5cm]{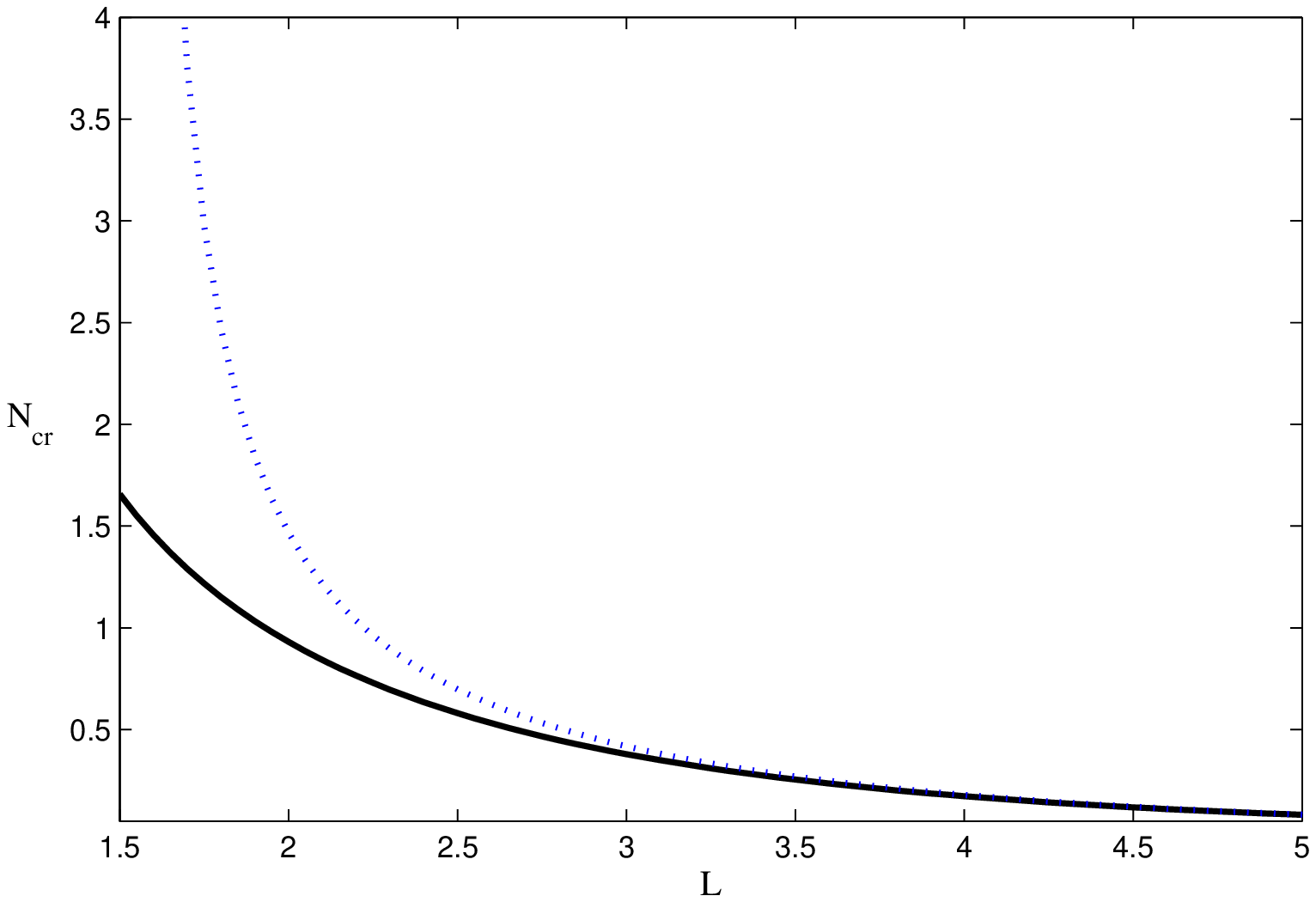} &
\includegraphics[height=5cm]{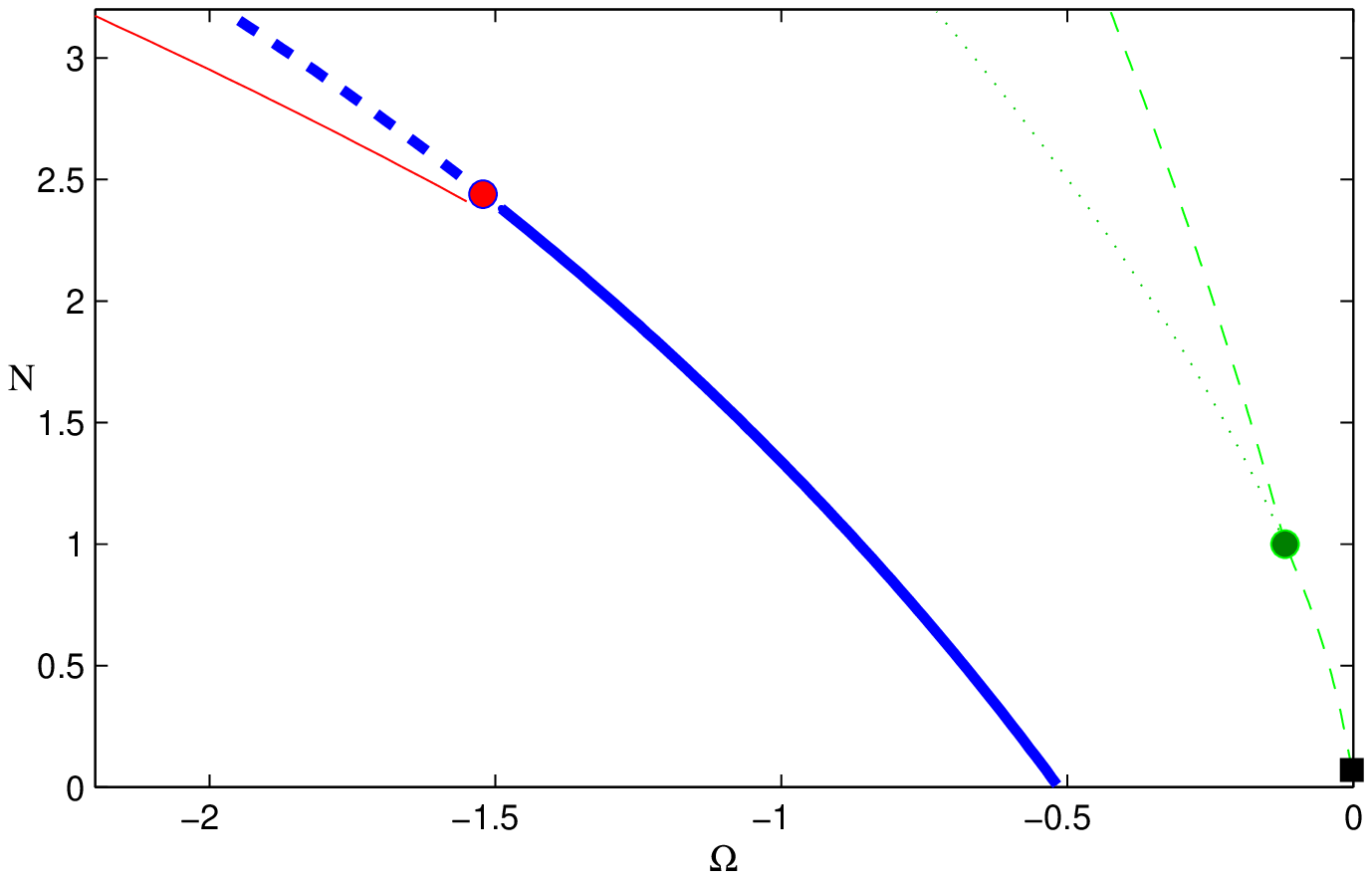} \\
\end{tabular}
\end{center}
\caption{(Color Online) The figure demonstrates the validity of $\cN_{cr}^{(0)}(L)$ as an approximation to
$\cN_{cr}(L)$. Panel (a) compares the linear finite dimensional estimation for the bifurcation point
$\cN^{(0)}_{cr}(L)$ and the actual numerical bifurcation point $\cN_{cr}$. The computations are for the double well
potential (\ref{poten}) $V_0=-1$ and $s= \frac{1}{4}$ and cubic nonlinearity. The curve $\cN_{cr}(L)$ is marked by a
solid (black) line and the curve $\cN^{(0)}_{cr}(L)$ is marked by a dotted (blue) line. Panel (b) shows a numerical
bifurcation diagram for the double well potential (\ref{poten}) $V_0=-1$, $s= \frac{1}{4}$ and $L=1.3$. The bifurcation
point $\cN_{cr}$ is marked by a (red) circle. For $\cN < \cN_{cr}$ the ground state marked by a thick (blue) solid line
is stable. For $\cN > \cN_{cr}$ the ground state is unstable and marked by a thick (blue) dashed line. The stable
asymmetric state which appears for $\cN > \cN_{cr}$ is marked by a thin (red) solid line. The unstable antisymmetric
state $(\Omega^{\cN}_1)$ is marked by a thin (light green) dashed line. The point $\cN$ for which the antisymmetric
state appears in the discrete spectrum is marked by a (black) square. Notice that in this bifurcation diagram there is
also a bifurcation from the antisymmetric branch. The state which bifurcates from the antisymmetric state is marked by
a (dark green) thin dotted line.} \label{fig:NcrL}
\end{figure}

We now investigate the limits of validity of $\cN_{cr}^{(0)}(L)$ as an approximation to $\cN_{cr}(L)$ by varying the
distance $L$ between the potential wells (\ref{poten}). For $L$ large, $\cN_{cr}^{(0)}$, given by equation
(\ref{Ncrit}), is close to the actual $\cN_{cr}(L)$, the exact threshold. In this case the eigenvalues of
$-\D_x^2+V_L(x)$, $\Omega_0(L)$ and $\Omega_1(L)$,  are close to each other; see Remark \ref{rem:Llarge}. Therefore,
the bifurcation occurs for small $\cN$ and one is in the regime of validity of Theorem \ref{theo:symbrkbif}. In figure
\ref{fig:NcrL} we display a comparison between the estimate for $\cN_{cr}$ based on the finite dimensional truncation,
$\cN^{(0)}_{cr}$, and the actual $\cN_{cr}$. For large $L$ the two values are close to each other. As $L$ is decreased
the wells approach one another and eventually, at $L=0$, merge to form a single well potential. As $L$ is decreased,
the  eigenvalues of the linear bound states $\Omega_0(L)$ and $\Omega_1(L)$ move farther apart. For some value of $L$,
$L_d$,  the eigenvalue of the excited state, $\Omega_1(L)$,  merges at $\Omega=0$, into the continuous spectrum. For $L
< L_d$ the estimate $\cN^{(0)}_{cr}$ is not correct. In fact, $\cN_{cr}^{(0)}(L) \to \infty$, while the actual value of
$\cN_{cr}(L)$ appears to be remain finite.
In Figure \ref{fig:NcrL}a we observe that for $L < 2$, $\cN^{(0)}$ and $\cN_{cr}$ diverge from one another and
eventually the approximation $\cN^{(0)}_{cr}(L)$ tends to infinity, while the actual $\cN_{cr}(L)$ remains finite.
Moreover, in Figure \ref{fig:NcrL}b we show a bifurcation diagram for small $L$ in which the discrete (excited state)
eigenvalue of $-\D_x^2+V_L$, $\Omega_1$,  does not exist, and yet there exists a symmetry breaking point $\cN_{cr}$.

\bigskip

\nit{\bf More general nonlinearities}

\begin{figure}[ht]
\begin{center}
\includegraphics[height=6cm]{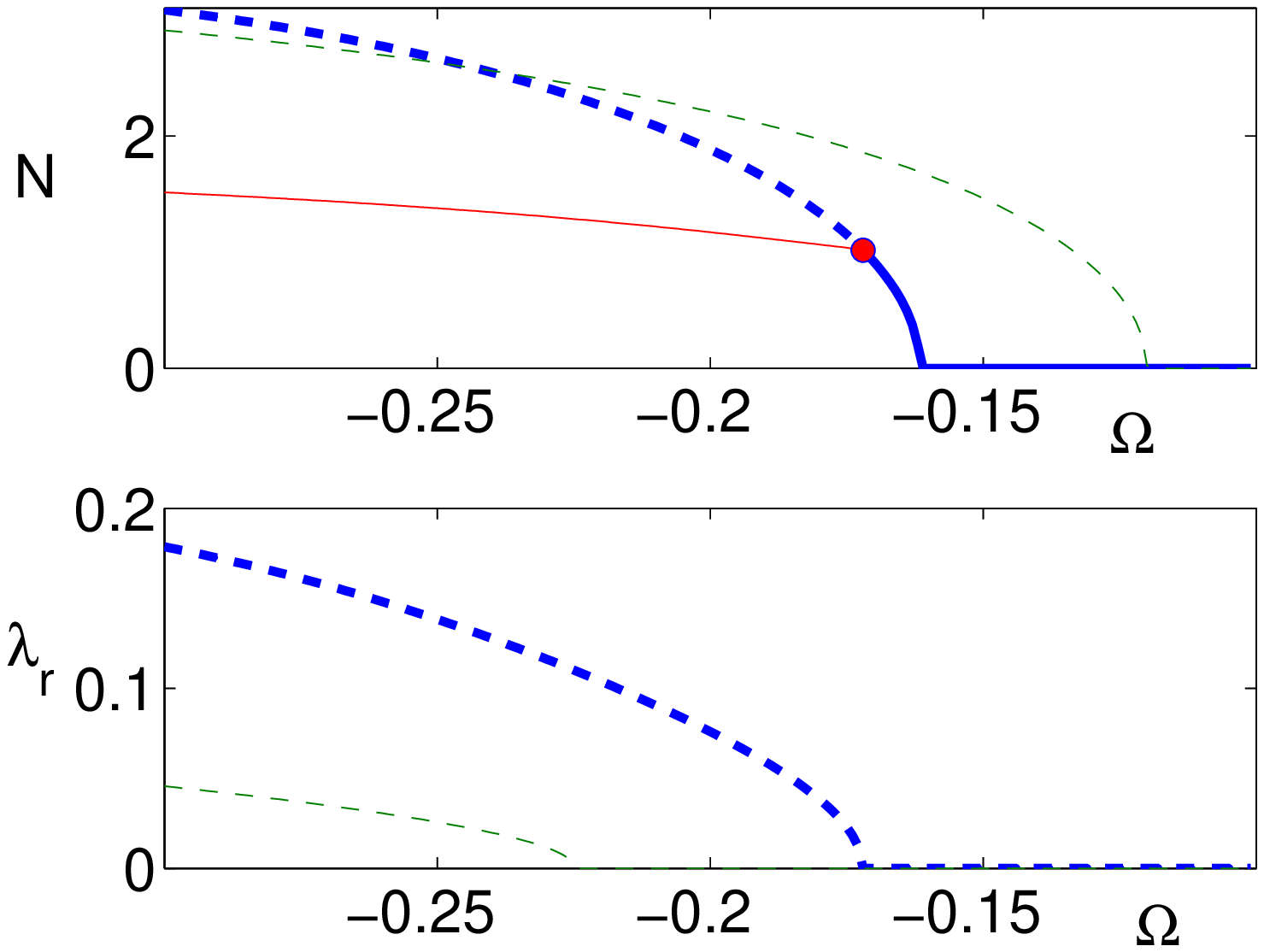}
\end{center}
\caption{Same as Figure \ref{bfig1} but for the
{\it quintic} nonlinearity. This serves to illustrate the analogies
between the bifurcation pictures but also their differences
(shifted critical point and also partial instability of the
anti-symmetric branch).}
\label{bfig2}
\end{figure}

To simplify the analysis, we assumed a cubic nonlinearity in NLS-GP. The analogue of  the  finite-dimensional approximation  (\ref{t3-eqns}) can be derived, for more general nonlinearities, by the same method. In this section we present numerical computations for general power law
nonlinearities such as $K[\psi \bar{\psi}]= (\psi \bar{\psi})^p$ and
 observe similar phenomena to the cubic case $p=1$.
 This is illustrated
e.g. in Figure \ref{bfig2}, presenting our numerical results
for the quintic case of $p=2$ (the relevant curves are analogous
to those of Figure \ref{bfig1}). It can be observed
that the higher order case possesses a similar bifurcation
diagram as the cubic case. However, the critical point for
the emergence of the asymmetric branch is now shifted to
$\Omega_{cr}^{(0)} \approx -0.1725$, i.e., considerably closer
to the linear limit. In fact, we have also examined the
septic case of $p=3$, finding that the relevant critical
point is further shifted in the latter to $\Omega_{cr}^{(0)} =-0.168$.
This can be easily understood as cases with higher $p$ are well-known
to be more prone to collapse-type instabilities (see e.g.
\cite{We:85}).  It may be an interesting separate venture
to identify $\Omega_{cr}^{(0)} $ as a function of $p$, and
possibly obtain a $p_{\star}$ such that $\forall \Omega<\Omega_0$,
the symmetric branch is unstable. We also note in passing that
bifurcation diagrams for higher values of $p$ may also bear
additional (to the shift in $\Omega_{cr}^{(0)} $) differences
from the cubic case; one such example in Figure \ref{bfig2}
is given by the presence of a linear instability (due to
a complex eigenvalue quartet emerging for $\Omega<-0.224$)
for the anti-symmetric branch. The latter was found to be linearly
stable in the cubic case of Fig. \ref{bfig1}.
\bigskip

\nit{\bf Nonlocal nonlinearities}

\begin{figure}[ht]
\begin{center}
\begin{tabular}{cc}
(a) & (b) \\
\includegraphics[height=5cm]{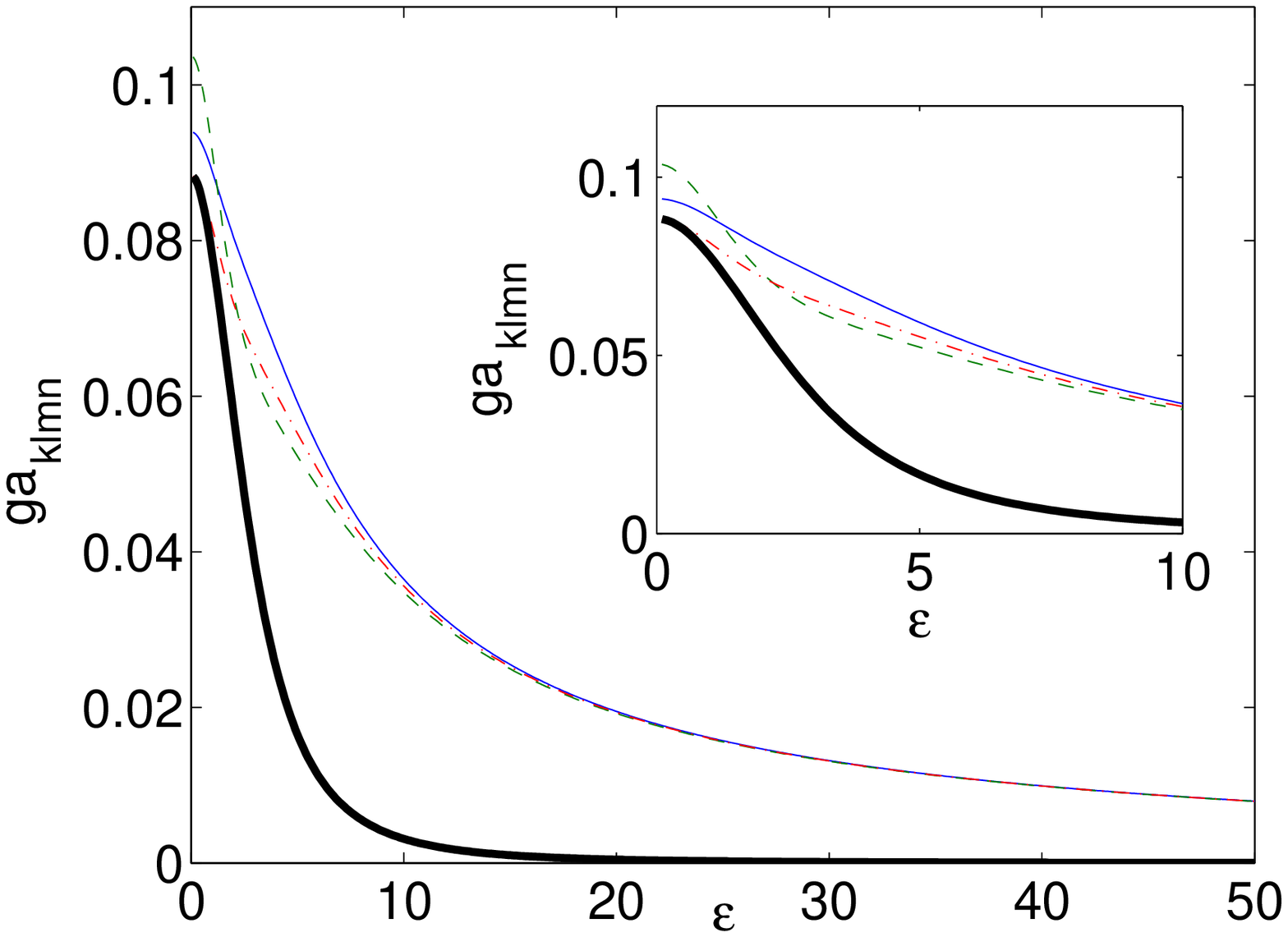} &
\includegraphics[height=5cm]{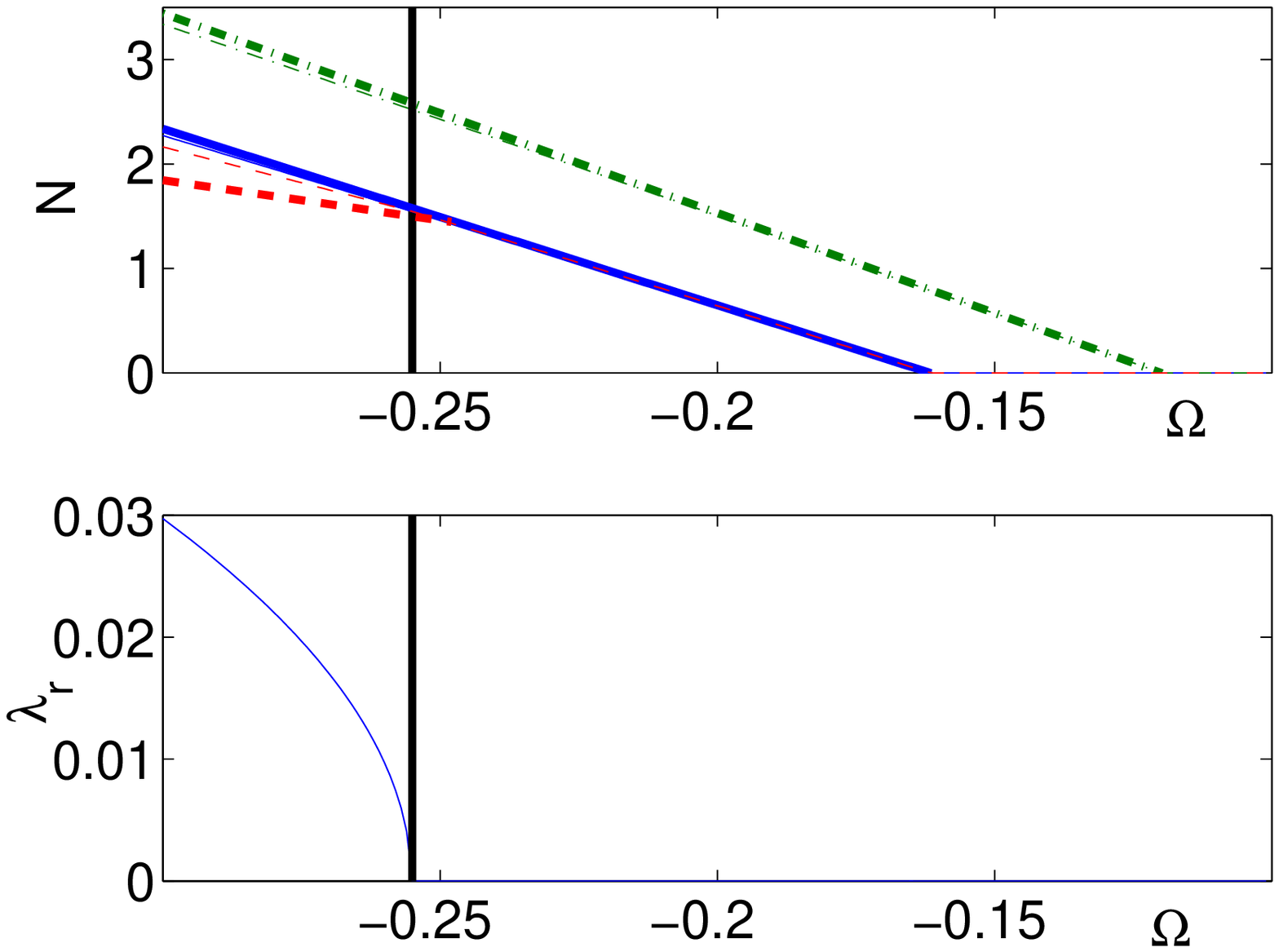} \\
\end{tabular}
\begin{tabular}{c}
(c) \\
\includegraphics[height=5cm]{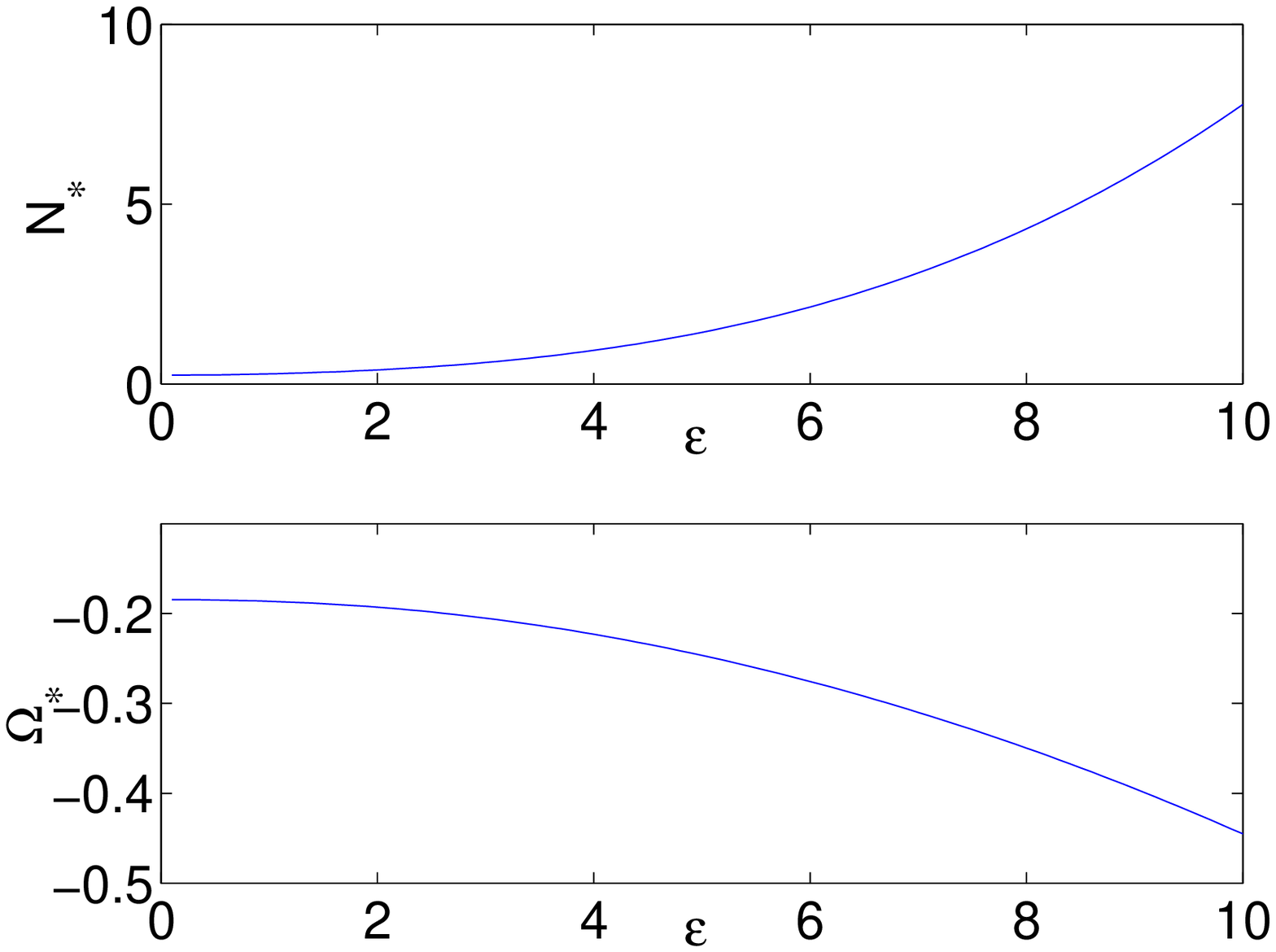} \\
\end{tabular}
\end{center}
\caption{This figure shows the nonlocal analog of Figure \ref{bfig1}.
Panel (a) shows the dependence of the (absolute value of the)
coefficients of the
finite-dimensional approximation on the non-locality parameter
$\epsilon$ ($\epsilon=0$ denotes the ``local'' nonlinearity limit).
The solid (blue) line denotes $a_{0000}$, the dashed (green) $a_{1111}$, the
dash-dotted (red) $a_{0110}$, while the thick solid (black) one denotes
$a_{0101}$. Panel (b) is analogous to panel (a) of Figure \ref{bfig1},
but now shown for the non-local case, with the non-locality parameter
$\epsilon=5$. Finally, panel (c) shows the dependence of the critical
point of the finite dimensional bifurcation $(N_{\star},\Omega_{\star})$,
on the non-locality parameter $\epsilon$.}
\label{bfig3}
\end{figure}

Finally, we consider the case of nonlocal nonlinearities, depending on a parameter $\epsilon$, the range of the nonlocal interaction.  In particular, consider  the case of a non-local nonlinearity of
the form:
\begin{eqnarray}
K[\psi \bar{\psi}]=\int_{-\infty}^{\infty} {\cal K}(x-y)
\psi(y) \bar{\psi}(y) dy,
\label{nonlocal}
\end{eqnarray}
where
\begin{eqnarray}
{\cal K}(x-y)=\frac{1}{2 \pi \epsilon^2} e^{-\frac{(x-y)^2}{2 \epsilon^2}}.
\label{kernel_num}
\end{eqnarray}
Here,  $\epsilon>0$ is a parameter controlling
the range of the non-local interaction. As $\epsilon$ tends to $0$,
${\cal K}(x-y) \rightarrow \delta(x-y)$ and we recover  the ``local'' cubic limit.
limit.  The form of the finite dimensional reduction, (\ref{t3-eqns}), does
not change; the only modification is that the coefficients
$a_{klmn}$ are now functions of the range of the interaction
 $\epsilon$.
The dependence of the coefficients, $a_{klmn}$ on $\epsilon$ is  displayed in panel (a) of Fig. \ref{bfig3}. The solid (blue) line
shows $|a_{0000}|$, the dashed (green) one corresponds to
$|a_{1111}|$, the dash-dotted (red) one to $|a_{1001}|=|a_{0110}|$
(due to symmetry),
while the thick solid (black) one to $|a_{0101}|=|a_{0011}|=|a_{1010}|$.
Notice in the inset how the coefficients asymptote smoothly to their
``local'' limit. Additionally, note the expected asymptotic relation $a_{1001}=a_{0011}$.
Also note the significant (decaying) dependence of the relevant
coefficients on the range of the interaction. The nature of this
dependence  indicates that while the character
of the bifurcation may be the same as in the case of local nonlinearities, its details
(such as the location of the critical points) depend sensitively
on the range of the non-local interaction. This is illustrated
in panel (b) for the specific case of $\epsilon=5$. In this panel
(which is analogous to panel (a) of Figure \ref{bfig1}, but
for the non-local case) the critical point for  emergence of
the asymmetric branch/instability of the symmetric branch is
shifted to $\Omega_{cr}^{(0)}=-0.2466$ (and the corresponding
$\cN_{cr}=1.4353$) in comparison to the numerically obtained
value of $\Omega_{cr}^{(0)} \approx -0.256$; the relative error
in the identification of the critical point (by the finite-dimensional
reduction) is in this case
of the order of $3.7 \%$, which can be attributed to the more strongly
nonlinear (i.e., occurring for higher value of $\cN_{cr}^{(0)}$) nature
of the bifurcation. However, as the finite-dimensional approximation
still yields a reliable estimate for the location of the critical
point, in panel (c) we use it to obtain an approximation to the
location of the critical point $(\Omega_{cr}^{(0)},\cN_{cr}^{(0)})$ as a function
of the non-locality parameter $\epsilon$.

 \section{Concluding remarks}
 We have studied the spontaneous symmetry breaking for a large class of NLS-GP equations, with double-well potentials.  Our analysis of the symmetry breaking bifurcation and the exchange
  of stability is based on an expansion, which to leading order in amplitude, is a superposition of a symmetric - antisymmetric pair of eigenstates of the linear Hamiltonian, $H$, whose energies are separated (gap condition (\ref{gapcond})\ ) from all other spectra of $H$. This gap condition holds for sufficiently large $L$ but breaks down as $L$ decreases. Nevertheless, numerical studies show the existence of a finite threshold for symmetry breaking. A theory  encompassing this phenomenon is of interest and is currently under investigation.

  \section{Appendix\ -\ Double wells}\label{sec:doublewells}
 In this discussion, we are going to follow the analysis of \cite{Harrell}.
Consider a (single well) real valued potential $v_0(x)$ on
$\mathbb{R}^n$ such that $v_0(x)\in L^r+L^\infty_\varepsilon $ for all
$1\le r\le q$ where $q\ge\max (n/2,2)$ for  $n\not= 4,$ $q>2$ for
$n=4.$ Then, multiplication by $v_0$ is a compact operator from $H^2$
to $L^2$ and
 $$H_0=-\Delta +v_0(x)$$
is a self adjoint operator on $L^2$ with domain $H^2.$

Consider now the double well potential:
 $$V_L=T_Lv_0T_{-L}+RT_Lv_0T_{-L}R$$
where $T_L$ and $R$ are the unitary operators:
 \begin{eqnarray}
 T_Lg(x_1,x_2,\ldots , x_n)&=&g(x_1+L,x_2,\ldots ,x_n)\nn\\
 R g(x_1,x_2,\ldots , x_n)&=&g(-x_1,x_2,\ldots ,x_n)\nn
 \end{eqnarray}
and the self adjoint operator:
 $$H_L=-\Delta+V_L(x)$$

\begin{prop}\label{dw1} Assume that $\omega<0$ is a nondegenerate
eigenvalue of
$H_0$ separated from the rest of the spectrum of $H_0$ by  a
distance greater than $2d_*.$ Denote by $\psi_\omega$ its
corresponding e-vector, $\|\psi_\omega\|_{L^2}=1.$ Then there exists
$L_0>0$ such that for $L\ge L_0$ the following are true:
 \begin{itemize}
 \item[(i)] $H_L$ has exactly two eigenvalues $\Omega_0(L)$ and
 $\Omega_1(L)$ nearer to $\omega$ than $2d_*.$ Moreover
 $\lim_{L\rightarrow\infty}\Omega_j(L)=\omega,\ j=0,1.$
 \item[(ii)] One can choose the normalized eigenvectors $\psi_j(L),\
 \|\psi_j(L)\|_{L^2}=1,$ corresponding to the e-values $\Omega_j(L),\ j=0,1$ such that they
satisfy:
 $$\lim_{L\rightarrow\infty}\|\psi_j(L)-(T_L\psi_\omega+(-1)^jRT_L\psi_\omega)/\sqrt{2}\|_{H^2}=0,\
 j=0,1.$$
 \item[(iii)] If $P^L_j$ are the orthogonal projections in $L^2$
 onto $\psi_j(L),\ j=0,1$ and $\tilde P_L=Id-P^L_0-P^L_1$ then there
 exists $d>0$ independent of $L$ such that:
 $$\|(H_L-\Omega)^{-1}\tilde P_L\|_{L^2\mapsto H^2}\ge d,\quad {\rm for\
 all}\ L\ge L_0\ {\rm and}\ |\Omega-\omega\|\le d_*.$$
 \end{itemize}
\end{prop}

{\it Proof:} For (i) we refer the reader to \cite{Harrell}. The
$L^2$ convergence in (ii) has also been proved there. The $H^2$
convergence follows from the following compactness argument. Let:
$$\psi^L_j=n_L\psi_j(L),\qquad j=0,1$$
where $n_L$ is such that $\|\psi^L_j\|_{H^2}=1,\ j=0,1.$ From the
eigenvector equations: $(H_L-\Omega(L))\psi^L=0,$ where we dropped the
index $j=0,1$ and the convergence $\Omega(L)\rightarrow\omega,$ see
part (i), we get
 \begin{equation}\label{l2lim}
 \lim_{L\rightarrow\infty}\|(-\Delta -\omega +V_L)\psi^L\|_{L^2}=0.
 \end{equation}
Denote:
 \begin{equation}\label{gLdef}g_L=(-\Delta -\omega)\psi^L\in L^2.\end{equation}
  Since $-\Delta-\omega:
H^2\mapsto L^2$ is bounded there exists a constant $C>0$ independent
of $L$ such that
$$\|g_L\|_{L^2}\le C.$$
Since $\omega<0,$ $-\Delta-\omega: H^2\mapsto L^2$ has a continuous
inverse then \eqref{l2lim} is equivalent to:
$$g_L+V_L(-\Delta-\omega)^{-1}g_L\rightarrow 0,\ {\rm in}\ L^2.$$
By expanding $V_L$ we get
 \begin{equation}\label{l2lim1}
 g_L+T_Lv_0(-\Delta-\omega)^{-1}T_{-L}g_L+RT_Lv_0(-\Delta-\omega)^{-1}T_{-L}Rg_L\rightarrow
 0.\end{equation}
 But $ v_0(-\Delta-\omega)^{-1}:L^2\mapsto L^2$ is
compact while the translation and reflection operators are unitary.
These and the uniform boundedness of $g_L$ lead to the existence of
$\psi\in L^2$ and $\tilde\psi\in L^2$ and a subsequence of $g_L,$
which we will redenote by $g_L,$ such that
\begin{equation}\label{l2lim2}\lim_{L\rightarrow\infty}
\|v_0(-\Delta -\omega)^{-1}T_{-L}g_L-\psi\|_{L^2}=0\ {\rm and}\
\lim_{L\rightarrow\infty}\|v_0(-\Delta -\omega
)^{-1}T_{-L}Rg_L-\tilde\psi\|_{L^2}=0.\end{equation} By plugging in
\eqref{l2lim1} and multiplying to the left by $T_{-L}$ we get
$$\lim_{L\rightarrow\infty}\|T_{-L}g_L+\psi
+RT_{2L}\tilde\psi\|_{L^2}=0.$$ But $RT_{2L}\tilde\psi$ converges
weakly to zero, hence $T_{-L}g_L$ converges weakly to $-\psi.$ By
plugging now in \eqref{l2lim2} and using compactness we get:
$$\psi+v_0(-\Delta-\omega)^{-1}\psi=0.$$
The latter shows that $(-\Delta-\omega)^{-1}\psi$ is an eigenvector of
$-\Delta+v_0$ corresponding to the eigenvalue $\omega.$ By nondegeneracy
of $\omega$ we get
 \begin{equation}\label{eqpsi}\psi=-n(-\Delta-\omega)\psi_\omega,\end{equation}
where $n$ is a constant. A similar argument shows
 \begin{equation}\label{eqtpsi}\tilde\psi=-\tilde
n(-\Delta-\omega)\psi_\omega,\end{equation} where $\tilde n$ is a
constant.

Combining \eqref{l2lim}-\eqref{eqtpsi} we get
 \begin{equation}
 \lim_{L\rightarrow\infty}\|(-\Delta-\omega)(\psi^L-nT_L\psi_\omega-\tilde
 nRT_L\psi_\omega)\|_{L^2}=0\end{equation}
which by the continuity of $(-\Delta-\omega)^{-1}:L^2\mapsto H^2$
implies
$$\lim_{L\rightarrow\infty}\|\psi^L-nT_L\psi_\omega-\tilde
 nRT_L\psi_\omega\|_{H^2}=0$$
Using now that $\|\psi^L\|_{H^2}=1$ and that the rescaled $\psi_j^L$
such that it has norm 1 in $L^2$ converges to
$(T_L\psi_\omega+(-1)^j
 RT_L\psi_\omega )/\sqrt{2}$ we get the conclusion of part (ii) for
 a subsequence first, then, by uniqueness of the limit, for all
 $L\rightarrow\infty.$

For part (iii), it suffices to show that there are no sequences
$(\Omega_L,\psi^L)\in [\omega-d_*,\omega+d_*]\times H^2$ with
$\|\psi^L\|_{H^2}=1$ and $\psi_L\bot\psi_j(L),\ j=0,1$ in $L^2$ such
that
 \begin{equation}\label{inv1}
 \lim_{L\rightarrow\infty}\|(H_L-\Omega_L)\psi^L\|_{L^2}=0.
 \end{equation}
The spectral estimate:
 $$\|(H_L-\Omega_L)\psi^L\|_{L^2}\ge  dist(\Omega_L,\sigma
 (H_L)\backslash\{\Omega_0(L),\Omega_1(L)\})\|\psi^L\|_{L^2}\ge
 d_*\|\psi^L\|_{L^2},$$
combined with \eqref{inv1} implies
 \begin{equation}\label{inv2}
 \lim_{L\rightarrow\infty}\|\psi^L\|_{L^2}=0.
 \end{equation}
In principle we can now employ the compactness argument in part (ii)
to get
\begin{equation}\label{inv3}
 \lim_{L\rightarrow\infty}\|\psi^L\|_{H^2}=0
 \end{equation}
which will contradict $\|\psi^L\|_{H^2}=1.$ More precisely,
\eqref{inv1}-\eqref{inv2} imply
 $$\lim_{L\rightarrow\infty}\|(-\Delta -\omega-d_*
 +V_L)\psi^L\|_{L^2}=0$$
 which, by repeating the argument after \eqref{l2lim} with $\omega $ replaced by $\omega+d_*$, gives
 $$\lim_{L\rightarrow\infty}\|\psi^L+T_L\psi_{\omega+d_*}+
 RT_L\tilde\psi_{\omega+d_*}\|_{H^2}=0$$
where $\psi_{\omega+d_*}$ and $\tilde\psi_{\omega+d_*}$ are
eigenvectors of $-\Delta+v_0$ corresponding to eigenvalue $\omega+d_*.$ But
the latter is not actually an eigenvalue, hence $\psi_{\omega+d_*}=0$
and $\tilde\psi_{\omega+d_*}=0.$ These show \eqref{inv3} and
finishes the proof of part (iii).

The proposition is now completely proven.

\begin{prop}\label{dw2}
\begin{align}
a_{1001}+2a_{1010}-a_{0000}\ &\le -\gamma< 0,\ \ \ {\rm and }\label{a-hyp1}\\
\frac{\Omega_1-\Omega_0}{|a_{1001}+2a_{1010}-a_{0000}|^2}\to0\ {\rm
as}\ L\uparrow\infty, &\label{ratio-cond1}
\end{align}
\end{prop}

These are now obvious from definition of $a_{ijkl}$, the continuity
of $N:H^2\times H^2\times H^2\mapsto L^2$ and the $H^2$ convergence
of $\psi_j(L)$ to the translations and reflections of the single
well eigenvector, see Proposition \ref{dw1} part (ii).

\bibliographystyle{plain}
\bibliography{sb-21-submit}
\end{document}

%% file: abbrevs.tex
\newcommand{\nc}{\newcommand}
\nc{\nit}{\noindent}
\nc{\nn}{\nonumber}
\nc{\D}{\partial}
\nc{\diff}[2]{\frac{d #1}{d #2}}
\nc{\diffn}[3]{\frac{d^{#3} #1}{d {#2}^{#3}}}
\nc{\pdiff}[2]{\frac{\partial #1}{\partial #2}}
\nc{\pdiffn}[3]{\frac{\partial^{#3} #1}{\partial{#2}^{#3}}}
\nc{\abs}[1] {\lvert #1 \rvert}
\nc{\cE}{{\cal E}}
\nc{\cH}{{\cal H}}
\nc{\cK}{{\cal K}}
\nc{\cN}{{\cal N}}
\nc{\Ncrz}{{{\cal N}_{cr}^{(0)}} }
\nc{\E}{{\Omega}}
\nc{\cF}{{\cal F}}
\nc{\cG}{{\cal G}}
\nc{\cV}{{\cal V}}
\nc{\Pzero}{{P_0}}
\nc{\Pone}{{P_1}}
\nc{\Pc}{{\tilde P}}
\nc{\vzero}{{\psi_0}}
\nc{\vone}{{\psi_1}}
\nc{\Ezero}{{\Omega_0}}
\nc{\Eone}{{\Omega_1}}
\nc{\cGin}{{\cal G}_{\rm in}}
\nc{\cGout}{{\cal G}_{\rm out}}
\nc{\cO}{{\cal O}}
\nc{\Lav}{{\cal L}_{\rm av}}
\nc{\cL}{{\cal L}}
\nc{\Ls}{{L_*}}
\nc{\cB}{{\cal B}}
\nc{\cQ}{{\cal Q}}
\nc{\cZ}{{\cal Z}}
\nc{\cR}{{\cal R}}
\nc{\cS}{{\cal S}}
\nc{\cT}{{\cal T}}
\nc{\cY}{{\cal Y}}
\nc{\vD}{{\vec D}}
\nc{\vE}{{\vec E}}
\nc{\vB}{{\vec B}}
\nc{\vH}{{\vec H}}
\nc{\ty}{{\tilde y}}
\nc{\tpsi}{{\Phi}}
\nc{\cP}{{\cal P}}
\nc{\tu}{{\tilde u}}
\nc{\tV}{{\tilde V}}
\nc{\bx}{{\bf x}}
\nc{\bk}{{\bf k}}
\nc{\bX}{{\bf X}}
\nc{\bXYZ}{{\bf XYZ}}
\nc{\bY}{{\bf Y}}
\nc{\bF}{{\bf F}}
\nc{\bS}{{\bf S}}
\nc{\dV}{{\delta V}}
\nc{\dv}{{\delta v}}
\nc{\dE}{{\delta E}}
\nc{\dPsi}{{\delta\Psi}}
\nc{\dph}{{\Delta\theta}}
\nc{\pp}{\perp}
\nc{\order}{{\cal O}}
\nc{\Eplus}{E_+}
\nc{\Eminus}{E_-}
\nc{\Epm}{E_\pm}
\nc{\Vav}{V_{\rm av}}
\nc{\Rin}{R_{\rm in}}
\nc{\Rout}{R_{\rm out}}
\nc{\eplus}{e_+}
\nc{\eminus}{e_-}
\nc{\epm}{e_\pm}
\nc{\eps}{\epsilon}
\nc{\half}{{1\over2}}
\nc{\veps}{\varepsilon}
\nc{\vnabla}{{\vec\nabla}}
\nc{\G}{\Gamma}
\nc{\w}{\omega}
\nc{\mh}{h}
\nc{\mg}{g}
\nc{\sgn}{{\rm sgn}}
\nc{\vphi}{\varphi}
\nc{\tlambda}{\tilde\lambda}
\nc{\be}{\begin{equation}}
\nc{\ee}{\end{equation}}
\nc{\ba}{\begin{eqnarray}}
\nc{\ea}{\end{eqnarray}}

\nc{\g}{\gamma}
\nc{\ol}{\overline}

\newtheorem{theo}{Theorem}[section]
\newtheorem{defin}{Definition}[section]
\newtheorem{prop}{Proposition}[section]

\newtheorem{cor}{Corollary}[section]
\newtheorem{rmk}{Remark}[section]
\newtheorem{example}{Example}[section]

\def\R{{\rm \rlap{I}\,\bf R}}

\nc{\pT}{\partial_T}
\nc{\pz}{\partial_z}
\nc{\pt}{\partial_t}
\nc{\la}{\langle}
\nc{\ra}{\rangle}
\nc{\infint}{\int_{-\infty}^{\infty}}
\nc{\halfwidth}{6.5cm}
\nc{\figwidth}{10cm}